%% file: mlst_2025.tex
\documentclass[12pt]{iopart}

\usepackage{graphicx}
\usepackage{hyperref}       
\usepackage{amssymb}
\usepackage{booktabs}
\usepackage{color}

\expandafter\let\csname equation*\endcsname\relax
\expandafter\let\csname endequation*\endcsname\relax

\usepackage{amsmath,amsfonts}
\usepackage{bm}
\usepackage[ruled,vlined,linesnumbered]{algorithm2e}
\SetKwInOut{Parameter}{parameter}

\SetCommentSty{mycommfont}

\input{preamble}

\begin{document}

\title[Transfer learning from first-principles calculations to experiments]{Transfer learning from first-principles calculations to experiments with chemistry-informed domain transformation}


\author{Yuta YAHAGI$^{1,2}$, Kiichi OBUCHI$^{1,2}$, 
Fumihiko KOSAKA$^{2}$ and Kota MATSUI$^{3}$}

\address{$^1$ NEC Corporation, Minato-ku, Tokyo, Japan, 108-8001}
\address{$^2$ National Institute of Advanced Industrial Science and Technology, Tsukuba, Japan, 305-8568}
\address{$^3$ Nagoya University, Nagoya, Japan, 466-8550}
\ead{yuta-yahagi@nec.com}
\vspace{10pt}
\begin{indented}
\item[]January 2025
\end{indented}

\begin{abstract}
  Simulation-to-Real (Sim2Real) transfer learning, the machine learning technique that efficiently solves a real-world task by leveraging knowledge from computational data, has received increasing attention in materials science as a promising solution to the scarcity of experimental data. 
  We proposed an efficient transfer learning scheme from first-principles calculations to experiments based on the chemistry-informed domain transformation that bridges the heterogeneous source and target domains by harnessing the underlying physics and chemistry. 
  The proposed method maps the computational data from the simulation space (source domain) into the space of  experimental data (target domain). 
  During this process, these qualitatively different domains are efficiently integrated by a couple of prior knowledge of chemistry, (1) the statistical ensemble, and (2) the relationship between source and target quantities.
  
  As a proof-of-concept, we predict the catalyst activity for the reverse water-gas shift reaction by using the abundant first-principles data in addition to the experimental data. 
  Through the demonstration, we confirmed that the transfer learning model exhibits positive transfer in accuracy and data efficiency.
  In particular, a significantly high accuracy was achieved despite using a few (less than ten) target data in domain transformation, whose accuracy is one order of magnitude smaller than that of a full scratch model trained with over 100 target data. 
  This result indicates that the proposed method leverages the high prediction performance with few target data, which helps to save the number of trials in real laboratories. 
\end{abstract}

\section{Introduction}
Machine learning (ML) is now a driving force for innovation in a broad range of industries fueled by the explosive growth of deep learning (DL) technologies. 
Beyond the traditional three paradigms of science, theory, experiment and computation, data science has emerged as a fourth paradigm of science~\cite{Hey2009-xt}. 
ML has paved the way not only for advanced information processing, but also for data-driven approaches in real-world challenges such as material development~\cite{Wei2019-fy,Zhou2019-dr, Schleder2019-os, Moosavi2020-kd}. 

In practice, the data-driven approach for materials often fails because of the lack of real data. 
Experimental data in materials science are scarce and non-scalable due to several reasons: (1) the high cost and time required for synthesis and measurement; (2) the disparate modality depending on measurement methods; and (3) the exploration bias, where the data collected is biased towards known or easily accessible regions of the material space, making ML tasks extrapolate beyond existing data. 
Although there are attempts to obtain large datasets by running high-throughput experiments~\cite{Nguyen2020-gh, Sugiyama2021-yl, Nakanowatari2021-wo,Wang2023-gu,Abed2024-mq}, combinatorial synthesis~\cite{Takeuchi2003-fh, Koinuma2004-qq, Uchida2018-hv, Iwasaki2020-mv,Ludwig2019-ds}, and laboratory automation~\cite{Shimizu2020-be,MacLeod2020-jx,Szymanski2023-yv}, the available materials and conditions are limited, and the data sizes are still insufficient, typically of the order of $O(100)$, below the requirements of common DL methods. 

On the other hand, computational data generated by first-principles calculations such as density functional theory (DFT) have been increasingly used in the context of ML. 
Compared to real experiments, numerical computations are scalable and easily automated, providing a large and abundant dataset. 
Such computational datasets obtained by high-throughput DFT calculations are available in a wide variety of materials~\cite{Curtarolo2013-av, Nishijima2014-sw, Iwasaki2021-pg, Chanussot2021-mq, Hayashi2022-ss, Tran2023-io, Zelezny2023-pd}. 
However, computational data also has limitations. 
Computer simulations rely on several approximations and assumptions, and thus systematic errors must appear. 
The impact of such errors cannot be known without comparison with experiments.
Although it is sometimes possible to reproduce a part of reality by performing high-fidelity calculations, such calculations are often expensive and significantly degrade their data throughput. 
Therefore, computational and experimental data are complementary to each other in which the former is lower cost but lower fidelity and the other is {\it vice versa}, motivating the introduction of transfer learning. 
In fact, a transfer learning framework, called the Sim2Real (Simulation-to-Real) transfer, has already been successfully introduced in some areas~\cite{kristinsson1992system,wang2018deep,bousmalis2018using,tobin2019real}.  

Although it is reasonable to hypothesize that the Sim2Real transfer is also effective for material discovery, this is still challenging due to the fundamental differences between first-principles calculations and real experiments. 
The most fatal issue is the difference in the scales: 
a first-principles calculation provides a microscopic description of a single (often simple) structure, which is represented by a set of atoms; an experimental measurement captures a macroscopic profile of a composite of various structures distributed near thermal equilibrium, which is usually difficult to control or even characterize. 

Kinetics is also a serious issue. 
To predict the kinetics of a real system, one needs to find the most plausible process among all combinations of possible elementary processes, which is explosive. 
Moreover, evaluation of a single elementary process is still expensive, as it requires a series of calculations to find a minimum energy path between an initial and a final state. 
For instance, let us consider the case of thermal catalysis on a solid-state surface~\cite{Motagamwala2021-iz}. 
A single first-principles calculation can provide a snapshot of an adsorption process on a simple periodic surface; 
it is far from a real experiment that measures a reaction rate resulting from the entire reaction path on a complex surface, involving various facets, surface reconstructions, and catalyst-support interactions. 

The central challenge addressed in this study is to bridge the gap between first-principles calculations and experiments, enabling Sim2Real transfer for material design. 
We propose a novel Sim2Real transfer method based on a chemistry-informed domain transformation that uses the laws of underlying physics and chemistry. 
The proposed method consists of two steps. 
It first transforms the domain of source computational data into that of target experimental data through formulas obtained in theoretical chemistry. 
Then, it solves the problem as homogeneous transfer learning, which can be easily solved with common transfer learning methods.
If the domain transformation is chemically appropriate and satisfies the transfer assumptions in the homogeneous domain~\cite{Shimodaira2000-ag,Quionero-Candela2008-vy}, a positive transfer can be expected in the second step. 
Ultimately, this approach allows us to build a predictive model that leverages the best of both worlds: the scalability and low cost of large-scale computational data while correcting for systematic errors using experimental data. 

This paper is organized as follows. 
Section 2 reviews related works in the field of Sim2Real transfer and transfer learning for materials. 
Section 3 describes the proposed method in detail. 
Section 4 demonstrates the effectiveness of the proposed method through a case study on catalyst activity prediction. 
Section 5 discusses the results and potential future directions. 
Finally, Section 6 concludes the paper with a summary of the findings and their implications.

In summary, the key contributions of this work are as follows:
\begin{itemize}
    \item \textbf{A novel Sim2Real transfer framework for materials}: We introduce a new transfer learning method using chemistry information that effectively tackles the challenges of data scarcity.
    \item \textbf{Practical application to catalyst discovery}: We predict catalyst activity using the proposed method, paving the way for practical applications.
    \item \textbf{Positive transfer}: Extensive experiments using real data confirmed the positive transfer improving accuracy and data efficiency, underscoring the effectiveness of our approach in practical scenarios.
\end{itemize}


%

\section{Related works}

\subsection{Simulation-to-Real (Sim2Real) Transfer}
Sim2Real transfer, the process of transferring knowledge from simulations to real-world applications, has been extensively studied in various fields such as computer vision, robotics, and some experimental sciences. 
Data generation and model training in simulation spaces can be conducted at a lower cost compared to real-world scenarios. 
However, there exists a gap between the simulator and the real world, which often serves as a major cause of failure when transferring data or models to real-world applications. 
This gap can arise, for instance, from inconsistencies in physical parameters (e.g., friction, damping, mass, density) or inaccuracies in physical modeling. 
Bridging the gap between simulation and the real world requires improving the simulation environment to better approximate reality. 
To achieve this, various approaches have been extensively studied, including system identification~\cite{kristinsson1992system}, which builds mathematical models of the physical system serving as the simulator; domain adaptation~\cite{wang2018deep, bousmalis2018using}, which adjusts the distribution of simulation data to align with real-world data; and domain randomization~\cite{tobin2019real}, which generates diverse pseudo-domains with randomized properties and trains models to perform robustly across all of them.
In this study, we do not focus on reconstructing the simulator itself, making system identification an unsuitable approach. 
Additionally, domain randomization requires performing simulations under numerous randomized settings, which would necessitate a large-scale computational environment when combined with high-cost simulations such as the DFT calculations targeted in this work. 
Therefore, we focus on domain adaptation, which can be achieved at a relatively lower computational cost.


\subsection{Transfer learning in Materials Science}
In the context of materials science, Sim2Real transfer is gaining traction as a promising approach to address the scarcity of experimental data.
For instance, Jha \textit{et al.}~\cite{Jha2019-zs} proposed a parameter transfer scheme from DFT datasets to a target dataset based on a neural network model. 
They successfully built a highly accurate model for predicting formation energy, which is significantly better than the predictive models built using only DFT datasets.
Wu \textit{et al.}~\cite{Wu2019-cp} proposed a Sim2Real transfer method to predict the thermal conductivity of polymers by transferring parameters learned in the source task predicting specific heat capacity with computational data. They demonstrated that the target task can benefit from the source task, achieving high accuracy with limited experimental data. 
Han and Choi also attempted a Sim2Real transfer to predict the NMR chemical shift from the computed magnetic shielding constant~\cite{Han2021-xd}.
Vermeire and Green applied a Sim2Real transfer approach to the solvation free energies leveraging a large computational dataset of solvent/solute combinations~\cite{Vermeire2021-wm}.
Aoki \textit{et al.}~\cite{Aoki2023-lg} developed a multitask learning framework of Flory--Huggins $\chi$ parameters for polymer-solvent systems. Their method integrates computational data with biased experimental data, enabling a highly generalized model to be applicable to a wider area than the coverage of experimental data. 

Data assimilation is a related approach to combine computational and experimental data.
Harashima \textit{et al.}~\cite{Harashima2021-od} proposed a data assimilation method in which a small number of experimental data and a large number of computational data are integrated. 
They applied this method to the search for permanent magnet compositions and succeeded in finding candidates that exhibit good properties at finite temperatures despite using few experimental data at finite temperatures with computational data limited to zero temperature.

In a similar sense to Sim2Real transfer, multi-fidelity transfer learning from abundant computational data to expensive but accurate ones is also a promising approach.
Smith \textit{et al.}~\cite{Smith2019-jk} provided highly accurate neural network potential by training a network to DFT data, then retraining to more accurate coupled cluster calculation data. 
Ju \textit{et al.}~\cite{Ju2021-iv} performed a feature-based transfer learning for predicting the lattice thermal conductivity of crystalline materials leveraging the scattering phase space data as a source data. 
Both properties are obtained from first-principles calculations, but the former is more expensive to compute than the latter. They demonstrated that feature transfer from the source task can significantly improve the prediction accuracy of the target task. 

Shim \textit{et al.} uses transfer learning to predict reaction conditions for new nucleophiles in Pd-catalyzed cross-coupling reactions. By combining transfer learning with active learning, the research efficiently identifies reaction conditions, successfully accelerating the discovery of chemical reactions~\cite{Shim2022-ir}.

Although there are many successful examples of transfer learning in materials science as listed above, these methods assume similar domains between the source and target where either feature space or target variable is identical; 
It is still challenging to integrate computational and experimental data across heterogeneous domains, which is the focus of this study.


\subsection{Physics/Chemistry-Informed Machine Learning}
Physics-informed machine learning integrates physical laws such as differential equations into machine learning models to improve their generalization and performance~\cite{Hao2022-pk}.
In material science, for example, Arora \textit{et al.}~\cite{Arora2022-yu} proposed a prediction model for the spatio-temporal evolution of deformation in elastic-viscoplastic materials by designing a physics-informed loss function, achieving high accuracy without increasing computational complexity. 

Similarly, chemistry-informed machine learning leverages chemical laws, which are mainly based on thermodynamics and statistical mechanics.
Bradford \textit{et al.}~\cite{Bradford2023-sv} developed a prediction model for ionic conductivity in polymer electrolytes based on a chemistry-informed neural network. Their architecture incorporates the Arrhenius equation in the final layers, which describes the temperature-activated processes, leading to a significant improvement. 
Ballard \textit{et al.}~\cite{Ballard2024-yq} performed machine learning modeling for polymerization processes with a chemistry-informed neural network. They incorporate the knowledge of kinetic models into the loss function, resulting in an improved model that outperforms conventional ones. 

This study, named chemistry-informed domain transformation, can also be regarded as a form of physics/chemistry-informed machine learning. 
While these previous studies incorporate the knowledge of physical and chemical laws into the models directly, our method uses it to establish some relation between heterogeneous domains for transfer learning, which is a unique challenge in this field.

\section{Method: Chemistry-informed domain transformation}
\label{sec:cidt}

\begin{figure}[ht]
  \centering
  \includegraphics[width=\columnwidth]{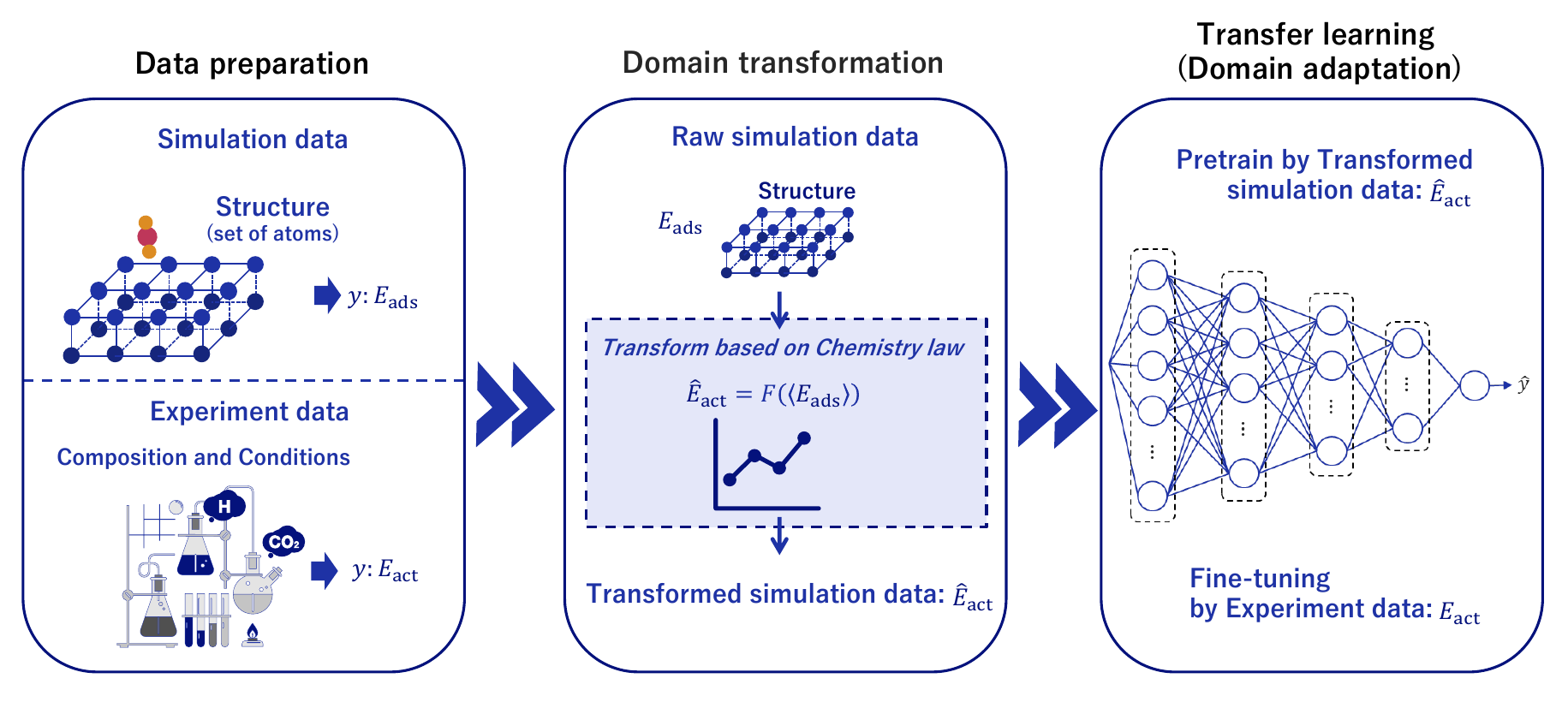}
  \caption{Schematics of our Sim2Real transfer learning framework for materials. Here, as an example, the adsorption energy $\Eads$ and activation energy $\Eact$ are assigned as computational and experimental quantities, respectively. See main text for further explanation.}
  \label{fig:schematics}
\end{figure}

\subsection{Overview}
The proposed method consists of two stages, a domain transformation part and a transfer learning part. 
Since the latter part is a standard set-up on a homogeneous domain adaptation, we focus on the former part. 

Let us begin with an overview of the process flow, schematically illustrated in Figure~\ref{fig:schematics}.
Our method takes as input two distinct datasets: a computational dataset composed of pairs of structures (set of atoms) and computed values (e.g. adsorption energy $\Eads$), and an experimental dataset consisting of pairs of experimental conditions (used in synthesis and measurement) and experimental values (e.g. activation energy $\Eact$). 
Experimental conditions are typically provided as numerical values; transfer learning requires the computational dataset to be converted into the identical format. 

To realize the domain transformation, this method leverages a couple of chemical information: the statistical ensemble under which the system follows, and the relational expression between the computational and experimental quantities, such as empirical rules or theoretical formulations, depicted as $F$. 

The domain transform part first establishes a correspondence between the structures and the conditions through evaluating the ensemble average, which allows us to aggregate all computational data under given conditions, forming the same feature space for the two domains. 
Then the (averaged) computational quantity is aligned with the experimental quantity with $F$. 
Eventually, it results in the transformed simulation data in the same domain as the experimental data, enabling homogeneous domain adaptation, a standard situation in transfer learning. 
We refer to this procedure as the “Chemistry-informed Domain Transform.” 

\begin{figure}[ht]
  \centering
  \includegraphics[width=\columnwidth]{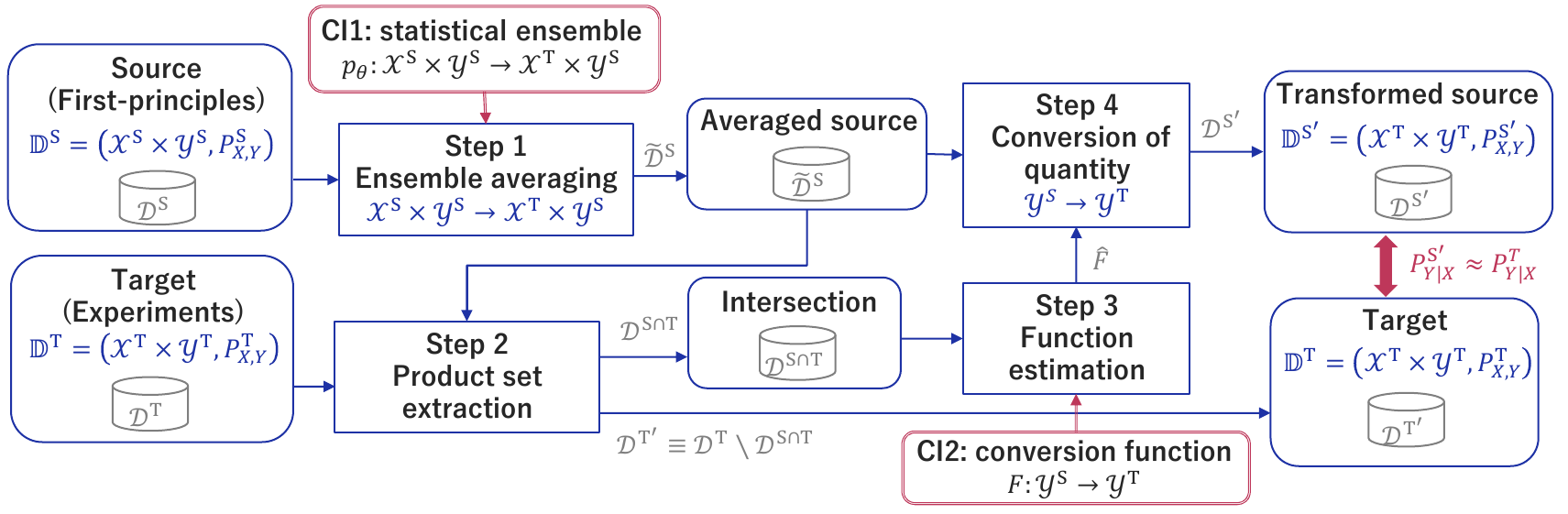}
  \caption{Schematics of the chemistry-informed domain transformation. The rectangle box represent the process, rounded rectangle represent the data and its domain, the red doubled-rounded rectangle represents the chemical information (CI), respectively.}
  \label{fig:transform}
\end{figure}

\subsection{Domain transformation procedure}
\label{sec:formalism}
We next show a detailed procedure of the chemistry-informed domain transformation as shown in Figure~\ref{fig:transform}. 
Let the domain be $\DD^{r}\equiv (\calX^{r} \times \calY^{r}, P^{r}_{X,Y})$ with
the feature space $\calX^{r}$, the output space $\calY^{r}$, and the joint probability distribution $P^{r}_{X,Y}$ where $r=\src$ or $\tgt$ represent the source and target domain, respectively.
Then the domain transformation aims to construct transformation rules for the source domain as follows:
\begin{equation}
\DD^\src=(\calX^\src\times\calY^\src, P^\src_{X,Y}) \to \DD^{\src'}=(\calX^\tgt\times\calY^\tgt, P^{\src'}_{X,Y}).   
\end{equation}
If such transformation rules are obtained, we can treat the source domain data (i.e., computational data) as target domain data (i.e., experimental data).

The source data $\calD^\src=\{(x^\src_i,y^\src_i)\}_{i=1}^{N_{\src}}$ is assumed to be obtained from first-principles calculations with the data size $N_{\src}$, where 
$x^\src = \{(\bm{r}_a; Z_a)\}_a^{\Natm} \in \calX^\src$ is a structure, a set of $\Natm$ atoms in certain positions $\bm{r}$ with elemental labels $Z \in \mathrm{Periodic Table}= \mathrm{\{H, He, Li,...\}}$. 
Furthermore, $y^\src \in \calY^\src$ is a physical quantity accessible from the calculations and correlated with the target measurement. 
The target data $\calD^\tgt=\{(x_i^\tgt,y_i^\tgt)\}_{i=1}^{N_{\tgt}}$ are obtained from experiments, where $x^\tgt \in \calX^\tgt$ is an experimental condition, such as chemical composition, temperature, pressure, etc., and $y^\tgt \in \calY^\tgt$ is a physical quantity measured in the experiments. 
$N_{\tgt}$ is the data size of the target data, assumed to be $N_{\tgt} \ll N_{\src}$.

Here, we additionally introduce the chemistry information (CI):
\begin{itemize}
  \item[\textbf{CI.1}] \textbf{Statistical ensemble} $p_\theta$: a statistical ensemble under which the system follows.
  \item[\textbf{CI.2}] \textbf{Conversion function} $F$: a relationship between quantities as a form of $F:\calY^\src \to \calY^\tgt$.
\end{itemize}

The domain transformation is achieved by following 4 steps: 
\begin{enumerate}
  \item \textbf{Ensemble averaging}: ($\calX^\src \times \calY^\src \to \calX^\tgt \times \calY^\src$). 
  \item \textbf{Product set extraction}. 
  \item \textbf{Function estimation}.
  \item \textbf{Conversion of quantity}: ($\calY^{\src}\to\calY^{\tgt}$).
\end{enumerate}
These procedures result in $\DD^\src\to\DD^{\src'}\approx \DD^{\tgt}$, reducing the problem to a homogeneous domain shift. 
In particular, from the construction of the domain transformation, if the output space $\mathcal{Y}$ is perfectly aligned, it is expected that the relationship known as covariate shift~\cite{Shimodaira2000-ag} will hold between the transformed computational data and the experimental data. 
That is, we can expect that the relation $P^{\src'}_{Y|X} \approx P^\tgt_{Y|X}$ approximately holds. 
Covariate shift is a typical model for the discrepancy between different domains, and various correction methods have been proposed in the machine learning community~\cite{Sugiyama2012-jm,fang2020rethinking}. 
By utilizing these methods, it is possible to achieve effective learning of prediction models under the covariate shift.

While we assume a single source domain in the following discussion for simplicity, the same methodology can be applied to multiple source domains with extending $F:(\calY^\src)^d \to \calY^\tgt$ where $d$ is the number of sources. 
Furthermore, there are no specific constraints on the dimensions of $x^{\src}, x^{\tgt}$ and $y^{\src}$. 

\subsubsection{\textbf{Ensemble averaging}:}
We first transform the feature space as $\calX^\src \times \calY^\src \to \calX^\tgt \times \calY^\src$.
The objective of this step is to obtain projected source data in the target feature space, 
$\tilde{\calD}^\src=\{(x^{\src'},\tilde{y}^{\src})\}$ where $(x^{\src'},\tilde{y}^{\src})\in\calX^\tgt \times \calY^\src$.
Here, $x^{\src'} \in \calX^\tgt$ is an auxiliary variable with the same feature space as $x^\tgt$.
$\tilde{y}^{\src} \in \calY^\src$ is the ensemble average of $y^{\src}$ associated with $x^{\src'}$, represented by: 
\begin{equation} \label{eq:ensemble_averaging}
  \tilde{y}^{\src} \equiv \EE_{x^{\src}}[y^{\src}(x^{\src})|x^{\src'}] \simeq \sum_{x^{\src}} p_{\hypara}(x^{\src}|x^{\src'})\ y^{\src}(x^{\src}),
\end{equation}
where $p_{\hypara}(x^{\src}|x^{\src'})$ denotes an occurrence probability of a structure $x^{\src}$ under a condition $x^{\src'}$ with additional fixed conditions $\hypara$, 
that is identical to a statistical ensemble in the context of statistical mechanics. 

The form of $p_{\hypara}(x^{\src}|x^{\src'})$ can be immediately determined from the assumed situation. 
For example, in the case of varying the temperature, Eq.~(\ref{eq:ensemble_averaging}) becomes:
\begin{equation}
  p(x^{\src}|x^{\src'}) \propto \exp {\left[ -\frac{E(x^{\src})}{\kB x^{\src'}} \right]},
\end{equation}
where $\kB$ is the Boltzmann constant, $E(x^{\src})$ is the energy of $x^{\src}$. 
This is known as the canonical ensemble. 

As an advanced example, let us consider the case of varying the chemical composition under isothermal conditions. 
In this case, $x^{\src'}$ is a chemical composition and $\hypara=T$ is the fixed temperature, leading: 
\begin{equation} \label{eq:ensemble_averaging_composition}
  p_{T}(x^{\src}|x^{\src'}) \propto \delta(x^{\src} \in x^{\src'})\exp{\left[ -\frac{E(x^{\src})}{\kB T} \right]},
\end{equation}
where $\delta(x^{\src} \in x^{\src'})$ is a Kronecker delta function that is 1 if a structure $x^{\src}$ has the composition of $x^{\src'}$ and 0 otherwise. 

To evaluate Eq.~(\ref{eq:ensemble_averaging}), we need to sample $x^{\src}$ and compute $y^{\src}$ with $E(x^{\src})$. 
This is usually accomplished by running a molecular dynamics or a Monte Carlo simulation with a given $x^{\src'}$, 
but it is computationally expensive and sometimes infeasible. 
In practice, further approximations will be used depending on the available source dataset $\calD^{\src}$, as demonstrated in the following section.
Alternatively, $\tilde{y}^{\src}$ can be obtained directly with calculations of physical properties or its surrogate model, if available.

\subsubsection{\textbf{Product set extraction}:}

After ensemble averaging, we can now compare $\tilde{\calD}^\src=\{(x^{\src'},\tilde{y}^{\src})\}$ with $\calD^\tgt=\{(x^\tgt,y^\tgt)\}$.
As the next step, we arrange a product set of them with respect to $x^{\src'}$ and $x^\tgt$, reffered as an intersection dataset: 
\begin{equation}
  \calD^{\src\cap\tgt} \equiv \{ (\tilde{y}^{\src}_i, y_j^{\tgt})\ |\ x^{\src'}_i = x_j^{\tgt} \}_{(i,j)} \subset \calY^\src \times \calY^\tgt ,
\end{equation}
whose size is $N_{\src\cap\tgt} \le N_{\tgt}$. 
This dataset provides a direct comparison between the source and target quantity, and is used in the following step to determine a map of $\tilde{y}^{\src}$ onto $\calY^\tgt$. 

To avoid leakage in future steps, the residual target data 
\begin{equation}
\calD^{\tgt'}= \calD^{\tgt} \setminus \calD^{\src\cap\tgt} 
\equiv\{ (x^{\tgt}_i, y_i^{\tgt})\ |\ y^{\tgt}_i \notin \calD^{\src\cap\tgt} \}_{i},
\end{equation}
are kept separately. 


\subsubsection{\textbf{Function estimation}:}

In this work, we assume that there exists a conversion function $F:\calY^\src \to \calY^\tgt$ from a source quantity to a target one. 
This $F$ describes some relation between $\tilde{y}^{\src}$ and $y^{\tgt}$. 
It may be either a black box function, an analytic function, or a non-parametric function, depending on the problem. 
The goal of this step is to estimate optimal $\hat{F}$ by using $\calD^{\src\cap\tgt}$. 

This task can potentially be solved efficiently by leveraging prior knowledge of the underlying physical and chemical principles. 
In certain combinations of $\tilde{y}^{\src}$ and $y^{\tgt}$, the form of $F$ and its parameter ranges can be deduced by the relevant formulas, such as the theoretical equation, the phenomenological relation and the natural laws. 
Note that since the sample size $N_{\src\cap\tgt}$ is generally limited, estimating the transformation function $F$ using a complex model may result in unstable outcomes. 
In other words, it is important to use the simplest possible model consistent with physical and chemical principles.

As a practical example, we will show a case of catalyst reaction, where the source quantity is the adsorption energy, and the target quantity is the activation energy. 
In this case, based on the theoretical chemistry, we can deduce that $F$ is a linear scaling function, which is discussed in detail in the next section. 

In addition to domain transformation, another important role of conversion with $F$ is to correct for systematic errors introduced by the approximations used in the various steps, such as first-principles calculations and ensemble averaging.
These errors are finally absorbed into the parameters of $\hat{F}$ estimated from real data, complementing the fidelity of source data within the capacity of $F$.

\subsubsection{\textbf{Conversion of quantity}:}

Finally, we convert $\tilde{y}^{\src}\in \calY^{\src}$ to $y^{\src'} \in \calY^{\tgt}$ with the estimated function $\hat{F}$, resulting in the source data in a transformed domain $\DD^{\src'}=\{(\calX^{\tgt}, \calY^{\tgt}, P^{\src'}_{X,Y})\}$. 
Since $F(\tilde{y}_i^S)$ and $y_j^T$ follow the same distribution for $x_i^{S'} = x_j^T$ and true $F$, we can expect  the covariate shift $P^{\src'}_{Y|X}\approx P^{\tgt}_{Y|X}$ hold approximately if the domain transformation is successful. 
Now it is ready to perform the homogeneous domain adaptation from $\DD^{\src'}$ to $\DD^{\tgt'}$.

\section{Demonstration}

In this section, we demonstrate the effectiveness of the proposed method through a practical case study on catalyst activity prediction for the reverse water gas shift (RWGS) reaction~\cite{Daza2016-tn, Gonzalez-Castano2021-bk}. 
We first describe the background and significance of this task. 
Then, we present the datasets used in this demonstration, followed by designing the task-specific model employing hypotheses appropriate to this situation. 
Finally, we show the results and implications. 

\subsection{Catalyst activity prediction for RWGS}

Catalysts are substances that improve the rate of chemical reactions by mediating them, and they can promote chemical reactions permanently without consuming themselves. 
They are indispensable to the chemical industry because they enable the production of important chemical substances faster, in larger quantities, and with less energy. 
The reverse water gas shift (RWGS) reaction is a crucial process in the chemical industry, where CO$_2$ is converted to CO, an ingredient in a variety of chemical products~\cite{Daza2016-tn, Gonzalez-Castano2021-bk}.
The development of novel catalysts has long been an area of interest within the field of chemical engineering. 

One of the most widely discussed performance metrics is the activation energy, which is the height of the energy barrier that must be overcome for a chemical reaction to proceed. 
By developing catalysts with lower activation energies, it is possible to produce products with lower energy and higher efficiency. 
In particular, for thermochemical catalysts, the operating temperature of the reactor can be lowered, increasing the flexibility of the overall design of the catalytic process. 
In experimental measurements, the activation energy can be estimated from the formation rate of the product, which requires significant time and human resources.

On the other hand, from the computational perspective, the activation energy cannot be computed directly because it requires the knowledge of the entire reaction path, which is computationally unfeasible. 
Instead, first-principles calculations commonly provide the adsorption energy, the binding energy between an adsorbate and a catalyst surface, which is relatively easy to obtain from structural relaxation~\cite{Sholl2009-yf}. 
Although this quantity is closely related to the activation energy, they are qualitatively different because the adsorption process is only a part of the total reaction processes. 

There are two types of quantities, the activation energy that is difficult to obtain but directly related to the target task, and the adsorption energy that is easy to obtain but indirectly related to that.
Therefore, this task is a suitable candidate for applying Sim2Real transfer with our method. 

\subsection{Datasets}

\begin{figure}
    \centering
    \begin{minipage}[b]{0.48\columnwidth}
        \centering
        \includegraphics[width=0.9\columnwidth]{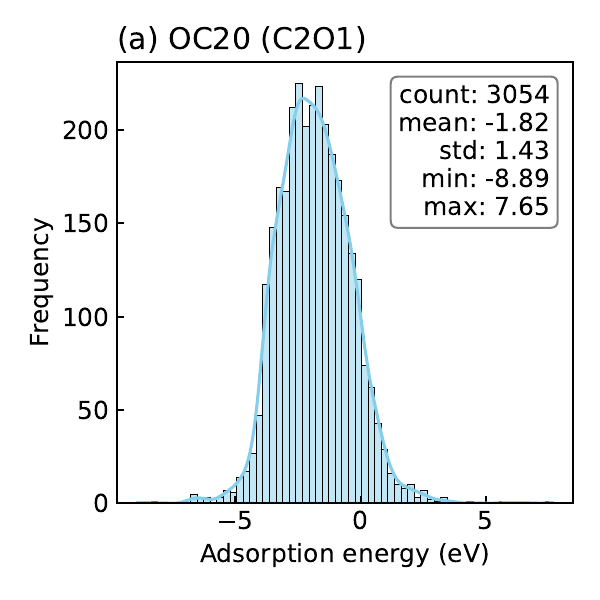}
    \end{minipage}
    \hspace{0.01\columnwidth}
    \begin{minipage}[b]{0.48\columnwidth}
        \centering
        \includegraphics[width=0.9\columnwidth]{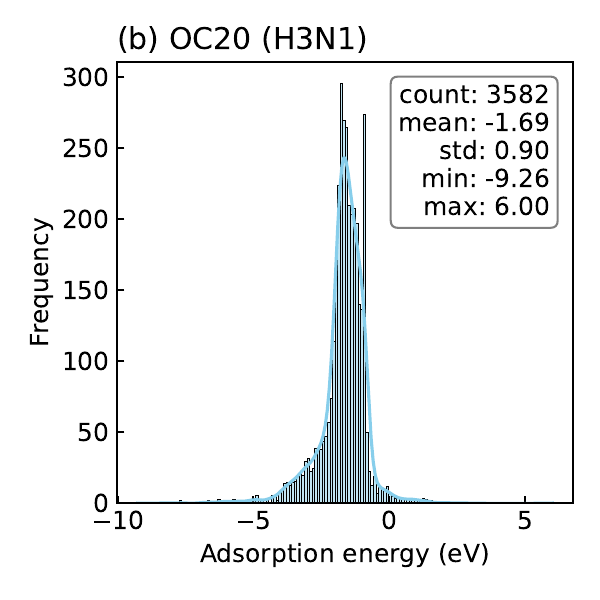}
    \end{minipage}
    \caption{Histograms of source data (\textit{OC20}) with (a) {\sf C2O1} and (b) {\sf H3N1}}
    \label{fig:SourceHistogram}
\end{figure}

Throughout this study, we use the \textit{Open Catalyst 2020 (OC20)} \cite{Chanussot2021-mq}, a simulation dataset based on density functional theory (DFT), as a source computational dataset.
It presents 48 datasets labeled with the adsorbate (=adsorption molecule) $M$; 
Each of them is a large collection of pairs of a slab structure $x^{\src(M)}$ and an adsorption energy $y^{\src_M}$ per adsorbate $M$.
Overall, it includes approximately $10^6$ entries of $\sim$ 2000 structures $\times$ 48 adsorbates. 
Figure~\ref{fig:SourceHistogram} plots histograms of $y^{\src}$ from $\calD^{\src(\mathsf{C2O1})}$ and $\calD^{\src(\mathsf{H3N1})}$ for example. 
Preparation details are explained in \ref{sec:app_oc20}


\begin{figure}
    \centering
    \begin{minipage}[b]{0.48\columnwidth}
        \centering
        \includegraphics[width=0.9\columnwidth]{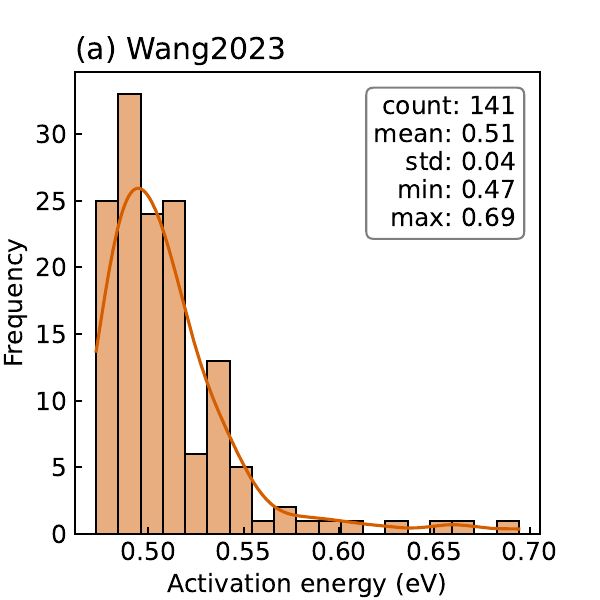}
    \end{minipage}
    \hspace{0.01\columnwidth}
    \begin{minipage}[b]{0.48\columnwidth}
        \centering
        \includegraphics[width=0.9\columnwidth]{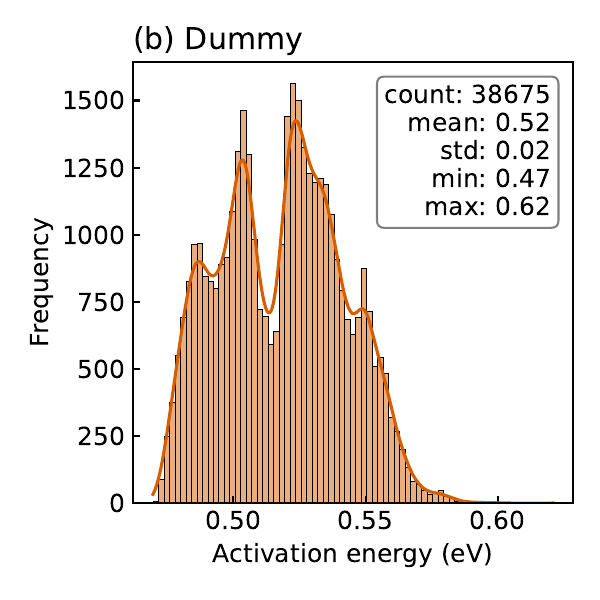}
    \end{minipage}
    \caption{Histograms of target data with (a) real experimental data (\textit{Wang2023}) and (b) the dummy data.}
    \label{fig:TargetHistogram}
\end{figure}

Regarding a target dataset, we refer to a high-throughput experimental dataset opened by Wang \textit{et al.} \cite{Wang2023-gu}, 
referred to \textit{Wang2023}.
Their dataset provides 300 pairs of a catalyst composition $x^{\tgt}$ and an activation energy $y^{\tgt}$. 
Here, we disregard the compositions including lanthanoids as they exceed the coverage of the \textit{OC20}, reducing the data size from 300 to 141. 
Its histogram is shown in Figure~\ref{fig:TargetHistogram}. The distribution is around the mean 0.510 with a standard deviation of 0.0359, but also has a tail in the high energy region over 0.55.
Preparation details are explained in \ref{sec:app_wang2023}

In addition, to verify the effectiveness of our method, we generated a dataset $\calD^{\dummy}$ with 38,674 pseudo-labeled data. 
It is prepared based on \textit{Wang2023}. 
Figure~\ref{fig:TargetHistogram} provides a histogram of the dummy data, showing a two-peak-like distribution between 0.470 and 0.621 with a standard deviation of 0.0233. 
The procedure for generating the dummy is given in the \ref{sec:app_dummy}. 

\subsection{Domain transformation}


We now proceed with domain transformation while specifying each step in Sec.~\ref{sec:formalism} for this catalyst activity prediction task. 
Details are explained as follows. 
Note here that \textit{OC20} is treated as a multiple source in this work. 

\subsubsection{\textbf{Ensemble averaging}: }
We first perform ensemble averaging for each $\calD^{\src(M)}$ to obtain $\tilde{\calD}^{\src(M)}=\{(x^{\src'},\tilde{y}^{\src(M)})\} \subset \calX^{\tgt}\times\calY^{\src}$. 
Formally, the statistical ensemble can be a probability distribution of surface adsorption structures $x^{\src}$ arising from a catalyst represented by the composition $x^{\src'}$ at an operating temperature $T$, like Eq.~(\ref{eq:ensemble_averaging_composition}). 

In this work, however, we approximately consider
\begin{equation}
  \label{eq:ensemble_averaging_adsorption}
  \tilde{y}^{\src(M)}(x^{\src'}) = \min \{ y^{\src(M)}_i | ~ x^{\src (M)}_i~\mathrm{has}~x^{\src'} \}_i,
\end{equation}
where "has" in the right-hand side means that a structure $x^{\src}_i$ has the composition of $x^{\src'}$. 
This is because the phase space of the adsorption structures is enormous to sample and thus requires simplification. 

Note that, Eq.~(\ref{eq:ensemble_averaging_adsorption}) corresponds to the low-temperature limit of Eq.~(\ref{eq:ensemble_averaging_composition});
In other words, this is nothing more than hypothesizing the following expression for the statistical ensemble:
\begin{equation}
   (\textbf{CI.1})~~~  p \propto \lim_{T\to +\infty} 
   \delta({x^{\src(M)}}\in x^{\src'(M)})
   ~\exp\left\{ {\frac{E(x^{\src(M)})}{\kB T}} \right\}.
\end{equation} 

Clearly, this is an overly bold approximation that deviates from the actual experimental conditions, 
and the sample size is insufficient for reliably exploring a minimal $y^{\src(M)}$, leading to significant systematic errors. 
However, these approximation errors may be partially mitigated in the subsequent function estimation step.

\subsubsection{\textbf{Product set extraction}:}
We then extract the product sets $\{\calD^{\src(M)\cap\tgt}\}$ for each $M$ from $\{\tilde{\calD}^{\src(M)}\}$ and $\calD^{\tgt}$. 
To measure the similarity between two compositions, we define: 
\begin{equation}
  d(x^A, x^B) = \sum_{Z=\mathrm{H, He, Li,...}}^{\mathrm{Periodic Table}} |c_Z(x^A) - c_Z(x^B)|,
\end{equation}
where $c_Z(x)$ is the concentration of element $Z$ in a composition $x$. 
Let $\threshold$ be a threshold, we obtain the product sets as 
\begin{equation}
  \calD^{\src(M)\cap\tgt} = \{ (\tilde{y}^{\src(M)}_i, y_j^{\tgt}) ~~~ | ~~~ d(x^{\src'}_i, x_j^{\tgt}) < \threshold \}_{(i,j)}.
\end{equation}
Here, we assume $\threshold=0.3$ for \textit{Wang2023}.

The product sets with the dummy dataset, $\calD^{\src(M)\cap \dummy}$, are obtained similarly, but assuming $\threshold=0.05$ due to the denseness of $\calD^{\dummy}$. 

To avoid data leakage, subtracted target datasets 
\begin{equation}
\calD^{\tgt'(M)}=\calD^{\tgt}\setminus \calD^{\src(M)\cap\tgt}
    \equiv
    \{ (x^{\tgt}_i, y_i^{\tgt})\ |\ y^{\tgt}_i \notin \calD^{\src(M)\cap\tgt} \}_{i},
\end{equation} 
are defined and used in the subsequent domain adaptation process. 
The number of $N_{\src(M)\cap T}$ depends on $M$, but is in the range of 10 to 13 for each $\calD^{\src(M)\cap \tgt}$. 
The precise values will be summarized later (See Table~\ref{tab:datasize_summary}).

\subsubsection{\textbf{Function estimation}:}
Next, we estimate a conversion function from adsorption energies to an activation energy, $F:(\calX^\src)^d\to\calY$, where it is a many-to-one mapping reflecting the multi-source situation. 
A form of $F$ can be deduced by referring to the empirical rule known as the Br\o nsted--Evans--Polanyi (BEP) relation~\cite{Norskov2002-up,Motagamwala2021-iz}:
\begin{enumerate}
    \item In many cases, catalytic activity is determined by a rate-determining step (RDS) in the reaction path. Thus, only limited adsorbates related to the RDS are sufficient for consideration. 
    \item There is a linear correlation between the activation energy and the adsorption energy on the RDS, 
\end{enumerate}
leading the following expression:
\begin{equation}
    (\textbf{CI.2})~~~F(\tilde{\bm{y}}^{\src}) = a_{M^*} \tilde{y}^{\src(M^*)} + b_{M^*},
\end{equation}
where $\tilde{\bm{y}}^{\src}\equiv \{ \tilde{y}^{\src(M)}\}_M$ is a vector obtained by concatenating $\tilde{y}^{\src(M)}$ over all sources, $M^*$ is an adsorbate selected among the sources. 
$a_{M^*}$ and $b_{M^*}$ are the coefficient and intercept of linear regression with $\calD^{\src(M^*)}$. 
Comparing the form of $F$ to the BEP relation, selecting $M^*$ corresponds to the first law:
Ideally, $M^*$ should be selected as the adsorbate involved in the RDS. 
The assumption of linear scaling immediately corresponds to the second law.

Estimation of $F$ is performed by multiple steps of source selection that determine $M^*$ and linear regression that determines $a_{M^*}$ and $b_{M^*}$, as presented in Algorithm~\ref{alg:estimateF}. 
While there are various methods for source selection, in this work we consider the two types of selection strategies: 
\begin{itemize}
    \item[\textbf{S1.}] \textbf{Prior knowledge of kinetics:} Selection of a source associated with any one of the reaction intermediates of RWGS.
    \item[\textbf{S2.}] \textbf{Regression performance to the BEP relation:}  Selection of a source that fit well with the activation energies in linear regression.
\end{itemize}

The strategy S1 treats $M^*$ as a sort of chemical information and selects it empirically. 
In the case of the RWGS reaction, the possible intermediates involved in RDS are $\chem{CO}, \chem{HCOO}$ and $\chem{COOH}$ \cite{Pahija2022-ih}; 
There is no exact match to these in the \textit{OC20}, the adsorbates listed in Table~\ref{tab:SelectedSource} are possible candidates for close matches. 

Contrary, the S2 strategy is a hybrid of chemistry-informed and data-driven; 
$M^*$ is selected from the sources that well describe the linear scaling. 
A source selection algorithm with this strategy is shown in Algorithm~\ref{alg:multiple-imputation}.
In order to perform source selection while dealing with data misalignment during integration, 
we adopt an algorithm combining the multiple imputation with variable selection, proposed by Wood, White, and Royston~\cite{Wood2008-vc}. 
Here, we refer to the occurrence frequency through iterations of multiple imputation as a criterion for source selection. 

\begin{algorithm}[ht]
  \label{alg:estimateF}
  \caption{Function estimation for $\hat{F}$.}
  \KwIn{$\{\calD^{\src(M) \cap \tgt}\}_M$: Intersection data, $F$: Conversion function}
  \KwOut{$\hat{F}$: Conversion function with estimated parameters.}
  \Begin{
    $M^*~\gets~\mathsf{SourceSelection}(\{\calD^{\src(M) \cap \tgt}\}_M)$\;
    $\hat{a}_{M^*}, \hat{b}_{M^*}~\gets~\mathsf{LinearRegression}(\calD^{\src(M^*) \cap \tgt})$\;
    $\hat{F}( \tilde{\bm{y}}^{ \src} )~\gets~\hat{a}_{M^*}\times \tilde{y}^{\src(M^*)} + \hat{b}_{M^*}$\;
  \Return $\hat{F}$\;
  }
\end{algorithm}

\begin{table}
    \caption{List of adsorbates (= label of sources) selected for each strategy.}
    \centering
    \begin{tabular}{l|c}
    \toprule
      Strategy & Selected adsorbates  \\
      \midrule
        S1 & {\sf C2O1, H1C2O1, H1C2O2, H1C2O2, H2C2O1, H2C2O2}   \\
      \midrule
      S2 (\textit{Wang2023}) & {\sf H3C1, H3C2, H5C2O2, H3C2O1, H3N1, H1O1, H1N1O1} \\
      \midrule
      S2 (dummy) & {\sf H5C2O1, N2O1, H3C2, H1C2O1, H4C2O2, C1, H2C1, H3C2O1, H3N1} \\
      \bottomrule
    \end{tabular}
    \label{tab:SelectedSource}
\end{table}

\begin{algorithm}[ht]
  \label{alg:multiple-imputation}
  \caption{Source selection with multiple-imputation method.}
  \KwIn{$\{\calD^{\src(M) \cap \tgt}\}_M$: Intersection data, $F$: Conversion function}
  \KwOut{$\bm{f}^{\mathrm{occ}}$: Occurrence frequency of each source among iteration.}
  \Parameter{$\TimeOfImputation$: Times of imputation iteration, $\LASSOlambda$: regularization coefficient.}
  \Begin{
  $\calD^{\mathrm{Join}} \gets \mathsf{OuterJoin}(\{\calD^{\src(M) \cap \tgt}\}_M)$\; 
  Standerize $\calD^{\mathrm{Join}}$\;
  $\bm{f}^{\mathrm{occ}}\gets \bm{0}$\;
  \For{$i=1$ to $\TimeOfImputation$}{
    $\calD^{\mathrm{Imputed}}\gets \calD^{\mathrm{Join}}$\;
    Missing data in $\calD^{\mathrm{Imputed}}\gets \mathcal{N}(0,1)$: Normal distribution\;
    $\hat{\bm{W}}^{\mathrm{LASSO}} \gets \mathsf{LASSO}(\calD^{\mathrm{Imputed}};\LASSOlambda)$\;
    \For{M}{
        \If{$\hat{W}^{\mathrm{LASSO}}_M \neq 0$}{$f_M^{\mathrm{occ}}\gets f_M^{\mathrm{occ}}+1$\;}
    }
  }
  \Return $\bm{f}^{\mathrm{occ}}$\;
  }
\end{algorithm}

\begin{figure}[ht]
\centering
    \centering
    \includegraphics[width=0.8\columnwidth]{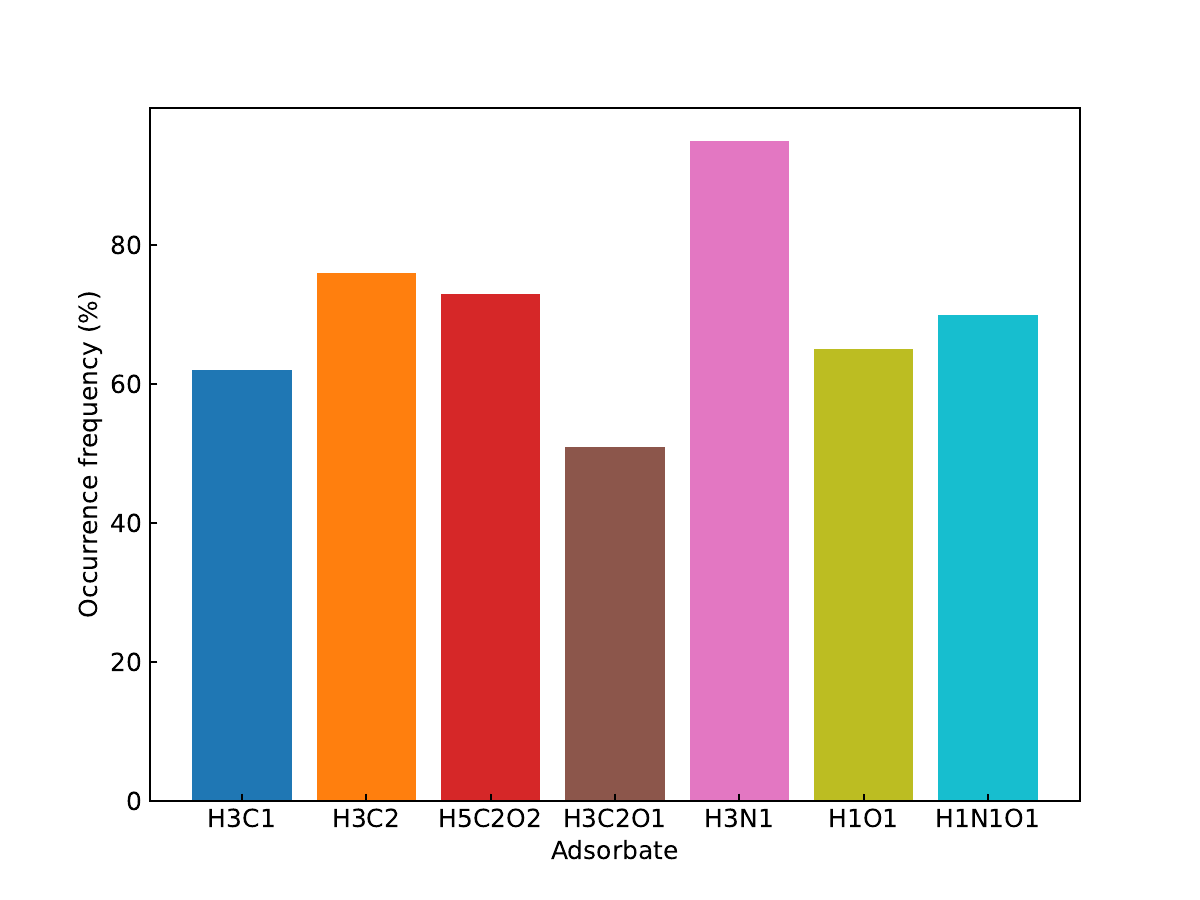}
    \caption{Occurrence frequency during iteration in the multiple imputation.}
    \label{fig:imputation_frequency}
\end{figure}
\begin{figure}[ht]
    \centering
    \includegraphics[width=0.5\columnwidth]{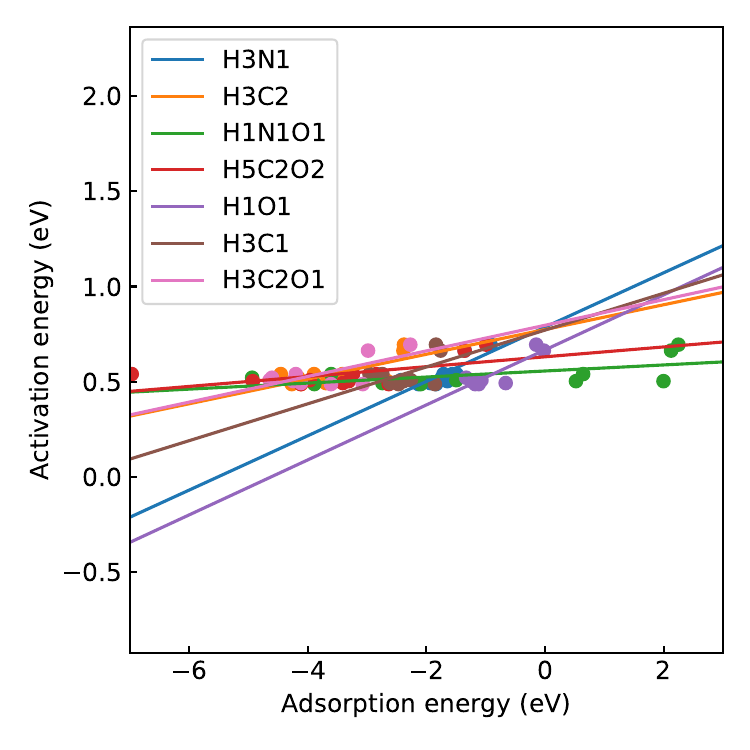}
    \caption{Linear regression of activation energy by adsorption energy of the sources selected with the S2 strategy.}
    \label{fig:regressionS2}
\end{figure}

Applying the Algorithm~\ref{alg:multiple-imputation} to $\{\calD^{\src(M)\cap \tgt}\}_M$, Figure~\ref{fig:imputation_frequency} displays the adsorbates whose occurrence frequency exceeds 50\%. Their linear regression lines are shown in Figure~\ref{fig:regressionS2}, presenting different lines depending on the sources. 
We consider these to be candidate sources and expect any one of them to give an adequate conversion. 
The selected sources from the strategy S2 are summarized in Table~\ref{tab:SelectedSource}. 
Detailed results and numerical values of the function estimation are given in \ref{sec:function_estimation}. 

Because the linear regression fits to the target data, some unanticipated effects and approximation errors may be mitigated within a capacity of $F$, thereby complementing the low fidelity of the simulations. 

Note that, while we rely on the linear scaling between activation and adsorption energy, this simple relation breaks down in certain cases, known as the Sabatier's principle~~\cite{Cheng2008-vn}. This principle states that the adsorption energy should not be too strong or too weak, since it follows a volcano curve. 
To capture this behavior, more representative functions such as a piecewise linear function can be used as the form of $F$, which is beyond the scope of this paper. 

\subsubsection{\textbf{Conversion of physical quantity}:}
\begin{figure}
    \centering
    \begin{minipage}[b]{0.48\columnwidth}
        \centering
        \includegraphics[width=0.9\columnwidth]{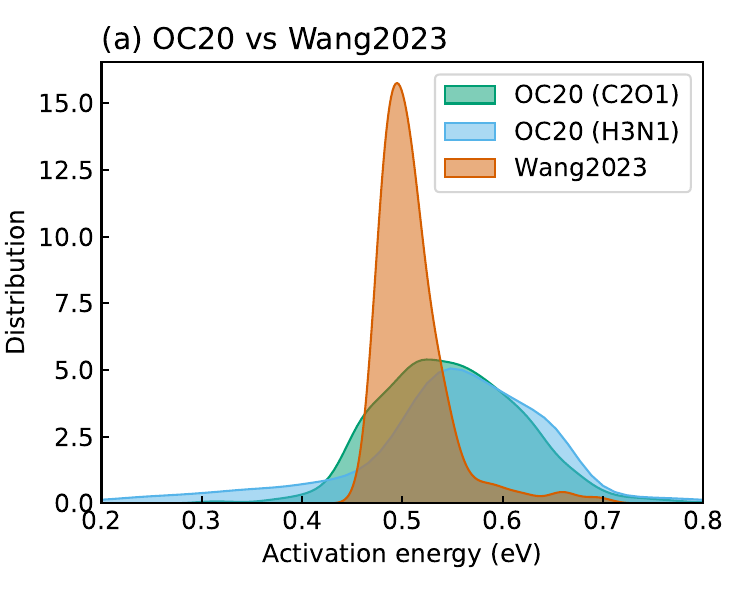}
    \end{minipage}
    \hspace{0.01\columnwidth}
    \begin{minipage}[b]{0.48\columnwidth}
        \centering
        \includegraphics[width=0.9\columnwidth]{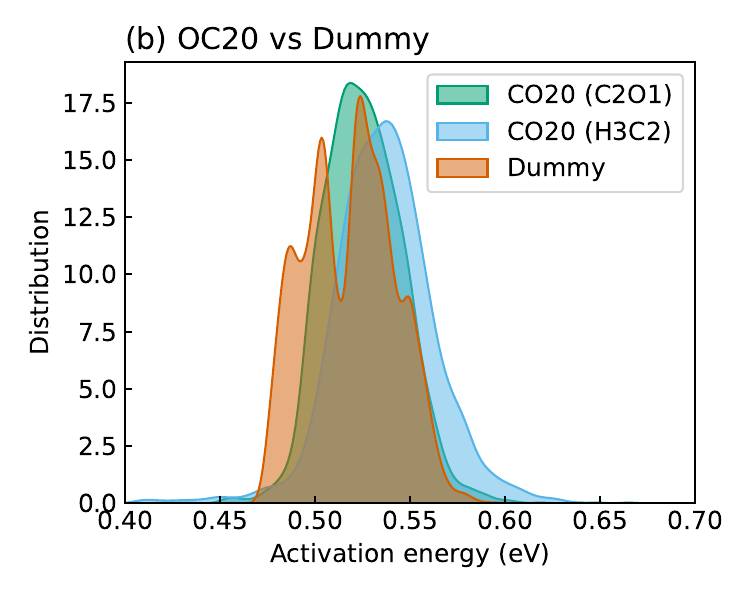}
    \end{minipage}
    \caption{Comparison of the converted source distributions and the target distribution of (a) \textit{Wang2023} and (b) dummy. The density distributions are estimated with a kernel density estimation method.}
    \label{fig:CompareDistro}
\end{figure}

We convert $\tilde{\bm{y}}_i^{\src}$ to ${y}^{\src'}$ with the estimated function $\hat{F}$, 
resulting in source data on the target space ${\calD}^{\src'}=\{({x}^{\src'}, {y}^{\src'})\}$, 
with the transformed source domain, $\DD^{\src'}=(\calX^\tgt\times \calY^\tgt, P_{X,Y}^{\src'})$. 
Figures~\ref{fig:CompareDistro} compare the label distribution of the transformed source data ($P(Y^{\src'})$) and target data ($P(Y^\tgt)$) for \textit{Wang2023} and the dummy data, respectively. 
In all adsorbates, the transformed source distribution is closer in shape to the target distribution than the raw \textit{OC20} distributions in Figures~\ref{fig:CompareDistro}. 
Furthermore, to quantitatively evaluate the effectiveness of our domain transformation, we calculated the discrepancy between distributions of the source and target domain before and after the transformation using the Wasserstein distance~\cite{peyre2019computational}. 
The Wasserstein distance can be interpreted as a measure of the cost required to transform one probability distribution into another, and a smaller Wasserstein distance suggests that the two distributions are closer to each other.
Table~\ref{tab:wasserstein} shows the Wasserstein distances between the source ({\sf C2O1}) distribution and the target distribution, suggesting that the distributions are closer to each other by domain transformation. 

\begin{table}
    \caption{Wasserstein distances between the raw or transformed source and target (\textit{Wang2023} or the dummy) distributions.}
    \centering
    \begin{tabular}{c|cc}
    \toprule
         & Raw source ({\sf C2O1}) & Transformed source ({\sf C2O1}) \\
    \midrule
    \textit{Wang2023} & 1.012 & 0.8727 \\
    Dummy & 1.043  & 0.2161 \\
    \bottomrule
    \end{tabular}
    \label{tab:wasserstein}
\end{table}

\subsubsection{\textbf{Homogeneous transfer learning}:}
\label{sec:model}

\begin{figure}
    \centering
    \includegraphics[width=0.8\linewidth]{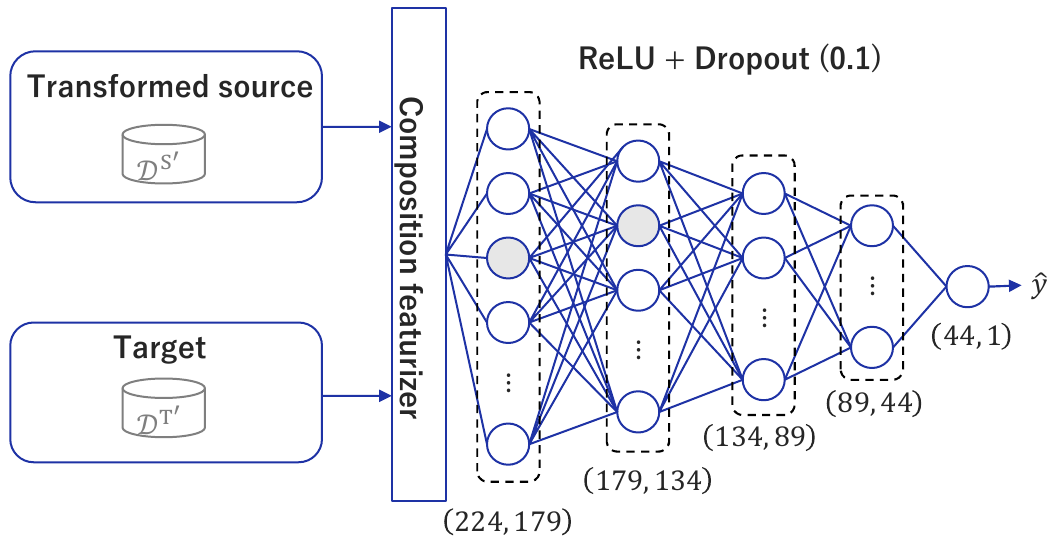}
    \caption{Architecture of the neural network prediction model.}
    \label{fig:NeuralNet}
\end{figure}

Finally, transfer learning pert performs the homogeneous domain adaptation. 
For input of the model, the composition is converted into the chemical descriptors generated by the {\sf XenonPy} code \cite{Liu2021-sp}.
We employ 224 descriptors that is relevant to our problem out of 290 descriptor, shown in \ref{sec:app_xenonpy}.
These descriptors are standardized using all compositions in \textit{OC20} dataset. 

The prediction model is based on a fully connected neural network (FCNN) model with five layers, as shown in Figure~\ref{fig:NeuralNet}. 
The number of units in each layer gradually decreases, starting from the input dimension 224, in the first layer, followed by 179, 134, 89, and 44, in the middle layers, with the final output layer having one unit. 
All units are activated by ReLU (rectified linear unit) and applied dropout at a rate of 0.1.
It should be noted that a similar five-layer FCNN architecture with compositional descriptors has already been successfully used in a previous study~\cite{Ju2021-iv}. 

To train a prediction model, we first prepare a source model from ${\calD'}^{\src}$.
With the target training, we compare two different methods: the fine-tuning (FT) that re-optimizes all layers, and the transfer learning (TL) that fixes the middle layers and retrain only the final layer.
During both training phases, the parameters are optimized so as to minimize the loss of mean squared error by the stochastic gradient descent algorithm with 32 batches with learning rate $=0.001$, which is implemented in the PyTorch framework~\cite{Paszke2019-zl}. 

\subsection{Training}
\label{sec:training}
We constructed a prediction model according to the method shown in Section \ref{sec:model}.
After domain transformation, the source model was trained using ${\calD'}^{\src}$, then the prediction models were prepared by retraining the source model on $\calD^{\tgt'}$. 
Here, we use 80\% of the target data for retraining and separate the remaining 20\% as the test data. 
Training for dummy target data proceeds similarly. 
More details of the training are presented in~\ref{sec:app_experiment}. 

\subsection{Performance evaluation}
\label{sec:results}
In this section, we present the evaluation results of the prediction model based on test error measurements, with a focus on learning efficiency. 
The performance evaluation is structured into two parts: validation using dummy data and practical assessment using real \textit{Wang2023} data. 

\begin{figure}[ht]
\centering
\includegraphics[width=0.7\columnwidth]{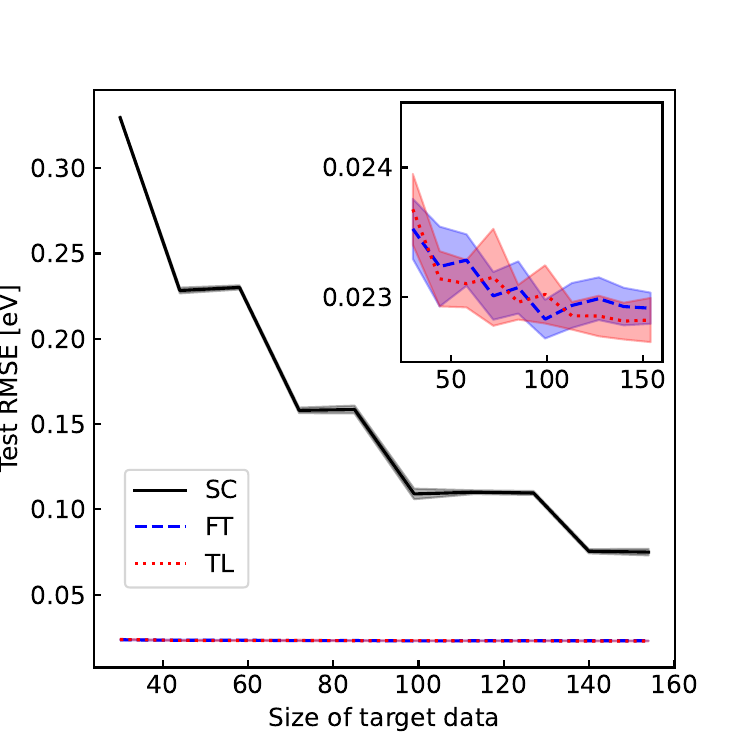}
\caption{Prediction accuracy measured by the test root mean squared error as a function of the size of used training data (a) with and (b) without the full-scratch model. The black solid line represents the full-scratch (SC) model, the blue dashed line the fine-tuning (FT), and the red dotted line the transfer-learning(TL), respectively (See main text). The hatched area represents the standard deviation in five trials.}
\label{fig:DummyDataSize}
\end{figure}

The effects of Sim2Real transfer were measured by evaluating the test loss as a function of the target training data size. 
Figure \ref{fig:DummyDataSize} shows the result of dummy target data pretrained with \textit{OC20} ({\sf C2O1}) dataset selected by S1 strategy. 
Here, we evaluate the mean loss and its standard deviation in five trials in which data were randomly selected from the training data set in each trial.

\subsubsection{Verification with dummy data: }
Compared to the case of the scratch model (trained with target data only), the transferred model clearly reduces the prediction error by approximately one order of magnitude. 
This is a direct evidence of positive transfer. 
Furthermore, the losses of transferred models have been minimized even without the target training steps, indicating that the source model has almost converged on optimal parameters in the target space. 
In comparison with the FT and the TL, there is little difference. 

\subsubsection{Demonstration with real data:}
\begin{figure}
    \centering
    \includegraphics[width=0.7\linewidth]{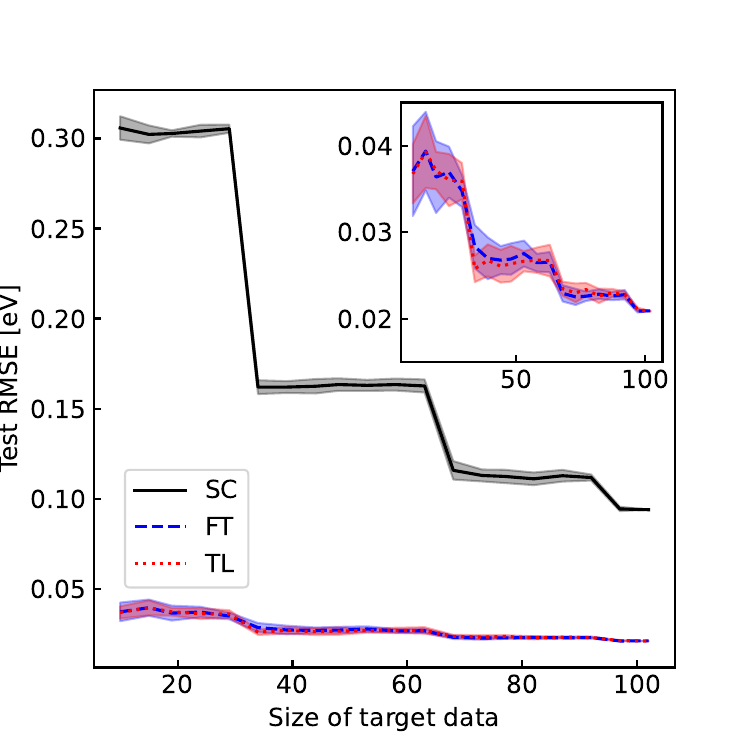}
    \caption{Test error as a function of the size of training target data similar to Figure~\ref{fig:DummyDataSize}.}
    \label{fig:DataSize}
\end{figure}

Next, let us evaluate performances for real experimental data from \textit{Wang2023} by using the pretrained model with \textit{OC20} ({\sf C2O1}) dataset. 
We also find positive transfer in this case from Figure~\ref{fig:DataSize}, indicating approximately 1/10 of the error reduction by knowledge transfer. 
Compared to Figure~\ref{fig:DummyDataSize}, the error at the start is higher, but still lower than any value of the loss of the scratch model; 
Thus, the transferred knowledge provides near-optimal parameters on the target domain. 
Similar to the previous section, the FT and the TL show almost the same performance. 

\subsubsection{Comparison of source datasets:}
At the end of this section, we will investigate source data dependencies. 
Since the transfer method does not affect the model performance as shown above, TL is used in this section. 

\begin{figure}
    \centering
    \includegraphics[width=0.6\linewidth]{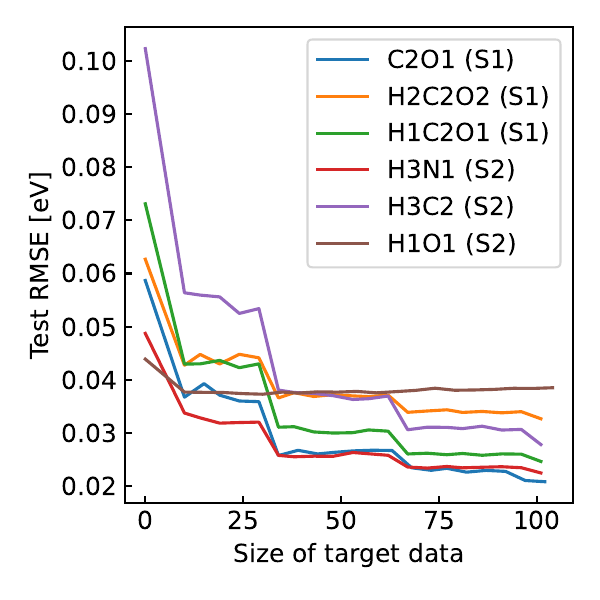}
    \caption{Test error as a function of the size of training target data with different sources.}
    \label{fig:DataSizeCompare}
\end{figure}

We evaluated the performances of the transferred model for different source datasets in \textit{OC20} selected with the strategy S1 or S2, summarized in Table~\ref{tab:datasize_summary}. 
Figure~\ref{fig:DataSizeCompare} shows the model performance for some of the sources included in the table. 
This result indicates that the final test errors depend on the source selection; however, the transferred model performs better than the scratch model regardless of the used source. 
Although {\sf C2O1} from the S1 strategy performs relatively better in terms of final accuracy, {\sf H3N1} from the S2 strategy reaches similar accuracy. 

It should be emphasized here that the frequency of occurrence from the S2 strategy (Figure~\ref{fig:imputation_frequency}) does not necessarily correspond to the performance of the transferred with the selected source. 
This is because the BEP relation claims that the activation energy can be linearly regressed on the adsorption energy of a molecule related to the RDS, but the converse is not true. 
Therefore, the source selection shown in Algorithm.~\ref{alg:multiple-imputation} is a procedure for exploring candidates rather than selecting the best one. 

\begin{table}
    \centering
    \caption{Summary of transfer learning with different source data.}
    \input{Tables/datasize_summary_S1S2}
    \label{tab:datasize_summary}
\end{table}

\subsection{Discussions}

Let us discuss the implications of the results and future perspectives.  

Our demonstrations in both a dummy dataset and a real dataset indicate that the proposed method successfully transfers the knowledge from source computational datasets beyond the domain gap, enabling highly efficient learning on the target real-world task. 
Pretraining with transformed source data, using only a few (about 10) target data for domain transformation, shows a significant improvement in prediction accuracy with reduced test losses by one order of magnitude. 

Moreover, the results show that the transferred knowledge gives nearly optimal parameters in the target model, suggesting a potential to reduce the number of experiments required to develop prediction models for novel catalysts. 


Although the current study demonstrates the potential of the proposed method, there are several areas for future improvement and exploration.

\begin{itemize}
  \item \textbf{Efficient sampling from ensemble:} With the source data preparation, the ensemble average is most expensive part and could be a bottleneck of this method. 
  For efficient data production, high-throughput computational frameworks for surface structures and molecular adsorption are available~\cite{Montoya2017-jx,Chanussot2021-mq}. 
  Alternatively, it would be a promising direction using a surrogate model for the adsorption energy~\cite{Lan2023-ng}. 
  \item \textbf{Expansion of chemical space coverage:} The obvious drawback of this approach is the need for diverse target data, even in small quantities. 
  Related to the previous point, it is required to define domain transformations that cover the global chemical space. 
  For example, evaluation of prediction uncertainty and Bayesian search for the next experiments could be used to expand the coverage efficiently. 
  \item \textbf{Optimal design of conversion function:} 
  The performance of the proposed method is highly dependent on the form of $F$ and the strategy of source selection. 
  Since there is a trade-off between function capacity and trainability, $F$ should be designed appropriately for the target task and the available data.
  Although this demonstration used a simple linear scaling function, it cannot work beyond the scope of the underlying Br\o nsted--Evans--Polanyi relation holding on a limited range of adsorption energy~\cite{Motagamwala2021-iz}. 
  Future work could explore more representative functions such as a piecewise linear function or a more complex volcano-like function to capture the global trend in catalysis known as the Sabatier principle~\cite{Cheng2008-vn}. 
  \item \textbf{Composing multiple sources:}
    In this study, one source data was selected and used for transition training. 
    However, in actual catalysis, activation energies and corresponding molecules are not uniquely determined either, since rate-limiting processes generally differ depending on the substance and surface~\cite{Pahija2022-ih}. 
    To deal with this situation, it is promising to use more flexible methods in function estimation or domain transformation, for instance, importance weighting per source or per instance~\cite{Sugiyama2012-jm}.
    \item \textbf{Refinement of the prediction model:} The current demonstration tested a simple parameter transfer framework for the prediction model. 
  As the next step, the model architecture and the training strategy could be further optimised for the target task. 
  Since any method that supports homogeneous domain adaptation can be applied here, we could explore not only different architectures but also different algorithms.
\end{itemize}

\section{Conclusions}
In this work, we proposed a simulation-to-real (Sim2Real) transfer learning method from first-principles data to experimental data based on the chemistry-informed domain transformation. 
The central problem addressed in this work is the domain difference between computational and experimental data, the former describes a microscopic picture of a material and the latter a macroscopic one.
The proposed method first transforms the source computational domain into the target experimental domain and then performs homogeneous domain adaptation to construct a prediction model for the target task. 
The domain transformation is achieved by the knowledge-based hypotheses on the statistical ensemble and the conversion function, which are designed to reflect the underlying chemical laws and enable linking qualitatively different features and physical quantities. 

As a proof of concept, we demonstrated the effectiveness of the proposed method in the prediction task of catalyst activity for the reverse water-gas shift reaction. 
Using the \textit{Open Catalyst 2020}~\cite{Chanussot2021-mq}, a large first-principles dataset, as the source dataset and the high-throughput experimental dataset from Wang \textit{et al.}~\cite{Wang2023-gu} as the target dataset, we construct a prediction model for the activation energy. 

Through a verification with dummy target data and a demonstration with the experimental data, we confirmed that the proposed method shows positive transfer of increasing accuracy and data efficiency compared to the model trained only with target data. 
The test losses of transferred model was approximately one order of magnitude smaller than that of full scratch model. 
Moreover, the source model has already been optimized in the target space, despite only using about 10 target data during the domain transformation. 

In conclusion, our proposed method helps to significantly reduce the number of experiments in real laboratories, leading to a drastic reduction of the cost and time of exploring novel materials. 
Finally, the essence of this approach is to integrate the four scientific paradigms, experiment, theory, computation, and data, thereby allowing full use of every possible knowledge. 
We believe that this work paves the way for the scheme of theory-informed transfer learning between computation and experiment, which should be effective for a variety of materials and tasks.

\section*{Acknowledgement}
We would thank Masahiko Ishida and Masato Hiroshima in NEC corporation, Japan for useful comments. We would also thank Yuya Ono in National Institute of Advanced Industrial Science and Technology, Japan for fruitful discussions.
This work was supported by Grants-in-Aid from the Japan Society for the Promotion of Science (JSPS) for Scientific Research (KAKENHI grant nos. 20K19871 and 24K20836 to K.M.).

\section*{References}
\bibliographystyle{unsrt}
\bibliography{mlst_2025}

\appendix

\section{Dataset preparation}
\label{sec:app_dataset}

In this work, we use the \textit{Open Catalyst 2020} (\textit{OC20}) \cite{Chanussot2021-mq} for a source computational dataset.
For a target experimental dataset, we use the \textit{Wang2023} data provided by Wang \textit{et al.} \cite{Wang2023-gu} or the dummy (pseudo-labeled) dataset generated from it. 

\subsection{Open Catalyst 2020}
\label{sec:app_oc20}
\textit{OC20} is a set of simulated data on molecular adsorption processes at metal surfaces, obtained by first-principles calculations based on the DFT. 
It randomly samples alloy surfaces within a low-index plane and adsorbs various molecules on them. 
While it consists of several tasks, we particularly focus on the initial structure to relaxed energy ({\sf IS2IR}) dataset because we are interested in the adsorption energy.
{\sf IS2IR} is a large collection of pairs of slab structures $S$ and adsorption energies $E^{M}$ per adsorbate $M$, including 
approximately $10^6$ entries ($\sim$10000 surfaces $\times$ 48 molecules).
We express these multiple source datasets as $\{ \calD^{\src (M)} \}_M^{48}$ with $\calD^{\src (M)} \equiv \{ (x^{\src (M)}_i, y_i^{\src (M)}) \}_{i=1}^{\sim10000}$.

\subsection{Wang2023}
\label{sec:app_wang2023}
Regarding a target dataset, we refer to a high-throughput experimental dataset opened by Wang \textit{et al.} \cite{Wang2023-gu}.
Their dataset presents measured data on the activity of RWGS on Pt-based catalysts synthesized on TiO2 supports.
The catalysts are prepared such that their compositions are $\chem{Pt}(3)$/$M1(l_1)$-$M2(l_2)$-$M3(l_3)$-$M4(l_4)$-$M5(l_5)$/$\chem{TiO_2}$ where $M_i$ is an element contained in a loading amount $l_i$ (wt\%) with the 3 wt\% $\chem{Pt}$, on the $\chem{TiO_2}$ support. 
This dataset consists of pairs of loading amount and $\chem{CO}$ formation rate, $r_{\chem{CO}}$, and is one of the largest experimental dataset of catalyst with a total of 300 entries (45 initial compositions and 255 compositions obtained by Bayesian search).
Here, we disregard compositions including lanthanoids as they exceed the coverage of \textit{OC20}, reducing the entries to 141 (38+103).

For simplicity, we assume that each catalyst is characterized by a chemical composition among $\{M_i\}$ plus $\chem{Pt}$ as $x^{\tgt}$.
We derive the activation energy $y^{\tgt}$ from the $\chem{CO}$ formation rate using the Arrhenius equation,
\begin{equation}
  r_{\chem{CO}}=A~\mathrm{exp}\left( \frac{y^{\tgt}}{\kB T} \right),
\end{equation}
where $T$ is the measurement temperature, $A$ is called the pre-exponential factor. $y^{\tgt}$ is measured in units of $\mathrm{eV}$.

In this work, $A$ is represented by the value of $\chem{Pt}(3)$/$\chem{Rb}(1)$-$\chem{Ba}(1)$-$\chem{Mo}(0.6)$-$\chem{Nb}(0.2)$/$\chem{TiO_2}$, irrespective of composition.
Eventually, we obtain the target dataset as $\calD^{\tgt} \equiv \{ (x_i^{\tgt}, y_i^{\tgt}) \}_{i=1}^{141}$.

\subsection{Dummy data}
\label{sec:app_dummy}
We generated a large dummy dataset $\calD^\dummy$ for Pt-containing quaternary alloys $\chem{Pt}({l_0})$–$M1({l_1})$–$M2({l_2})$–$M3({l_3})$ from \textit{Wang2023}.
To avoid combination explosions, we limit the constituent elements $M1, M2$, and $M3$ to those listed in the Table~\ref{tab:element-pool}.
In addition, we only consider the compositions with $l_i \le 3$. 

First, a prediction model $\hat{h}$ trained by \textit{Wang2023} is prepared. 
Because only $\calD^{\tgt}$ is not enough to train with a neural network-based model, we here employ $\hat{h}$ as a gradient boosting decision tree model implemented in {\sf LightGBM} code~\cite{Ke2017-hv}. 
To train the model, compositions are converted into the {\sf XenonPy} descriptor (See~\ref{sec:app_xenonpy}).
The essential parameters are listed in Table~\ref{tab:lgb-params}.

The dummy generation method is explained in Algorithm~\ref{alg:dummy-generation}. 
In this process, all possible compositions $\chem{Pt}(l_0)$-$M1(l_1)$-...-$M_{\numelem}(l_{\numelem})$ satisfying $l_i\le \maxindex$ are generated combinatorially.
Running it with $\numelem = 3$ and $\maxindex=3$, we finally obtain 38674 dummy data.

\begin{table}
    \caption{List of elements considered in the dummy generation.}
    \centering
    \begin{tabular}{c|c}
        \toprule
         & Elements \\
         \midrule
        Ordinary elements & Al, Si, Ga, Ge \\
        4d-transition metals & Nb, Mo, Tc, Ru, Rh, Pd \\
        5d-transition metals & Ta, W, Re, Os, Ir\\
        \bottomrule
    \end{tabular}
    \label{tab:element-pool}
\end{table}

\begin{table}
    \caption{List of parameters of the {\sf LightGBM} for generating pseudo-labels.}
    \centering
    \begin{tabular}{cc}
        \toprule
         Parameter & Value \\
         \midrule
        Data sample strategy & bagging \\
        Number of iterations & 100 \\
        Learning rate & 0.1 \\
        \bottomrule
    \end{tabular}
    \label{tab:lgb-params}
\end{table}

\begin{algorithm}[ht]
  \label{alg:dummy-generation}
  \caption{Process of the dummy data generation.}
  \KwIn{$\hat{h}$: Prediction model trained by original target data, $\mathcal{E}$: List of elements.}
  \KwOut{$\calD^{\dummy}$: Dummy data.}
  \Parameter{$\numelem$: Number of constituents, $\maxindex$: Maximum composition index}
  \Begin{
  $\calD^{\dummy} \gets \varnothing$\;
  \For{$\bm{M}=(M1,...,M{\numelem})$ in $\mathcal{E}^{\numelem}$}{
    \For{$\bm{l}=(l_0, l_1, ..., l_{\numelem})$ in $\{1,...,\maxindex\}^{\numelem + 1}$}{
        \If{$\mathsf{GratestCommonDevisor}(\bm{l})==1$}{ 
            $x_{\mathrm{new}}\gets \chem{Pt}(l_0)$-$M1(l_1)$-...-$M_{\numelem}(l_{\numelem})$\;
            $z_{\mathrm{new}} \gets \mathsf{CompositionalFeaturizer}(x_{\mathrm{new}})$
            $y_{\mathrm{new}}\gets \hat{h}(z_{\mathrm{new}})$\;
            Append $(x_{\mathrm{new}}, y_{\mathrm{new}})$ to $\calD^{\dummy}$\;
        }
    }
  }
  \Return $\calD^{\dummy}$\;
  }
\end{algorithm}

\section{Composition featurizer}
\label{sec:app_xenonpy}
The {\sf XenonPy} compositional featurizer is a tool within the {\sf XenonPy} library that converts raw chemical composition data into informative numerical descriptors suitable for machine learning~\cite{Liu2021-sp}. 
It can calculate 290 compositional features for a given chemical composition. 
This calculation uses the information of the 58 elemental properties. 
The compositional descriptors are obtained from the five calculations, {\sf weighted-sum, weighted-average, weighted-variance, max-pooling}, and {\sf min-pooling}, using the 58 elemental properties and the composition ratio of each element ($58\times 5 = 290$). 
For more details, see the code documentation (https://xenonpy.readthedocs.io/en/latest/features.html). 

In this work, since all $f$-block elements are excluded from our datasets in advance, we ignore 
{\sf num\_f\_unfilled} and {\sf num\_f\_valence} out of the 58 elemental properties. 
Moreover, we omit the {\sf weighted-sum} calculation as it is identical to the {\sf weighted-average} for a periodic system.
Consequently, we use $56 \times 4 = 224$ descriptors.

\section{Experimental details}
\label{sec:app_experiment}

\subsection{Function estimation}
\label{sec:function_estimation}

\begin{figure}
    \centering
    \includegraphics[width=0.7\linewidth]{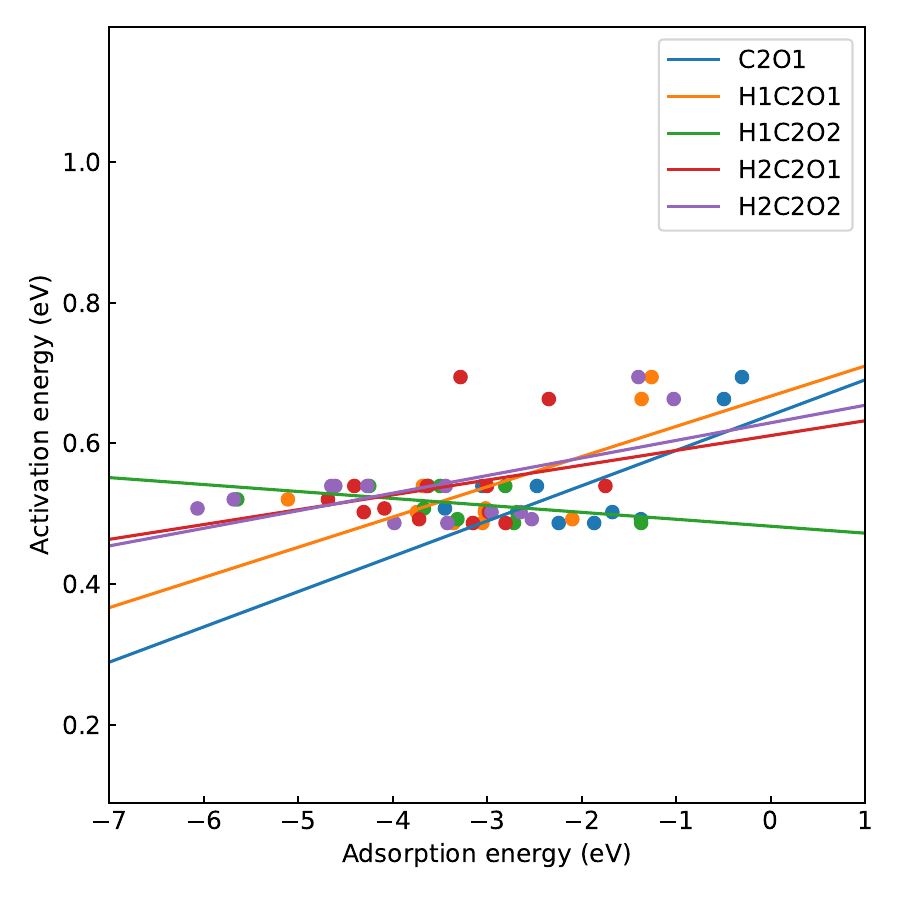}
    \caption{Linear scaling between activation energy and adsorption energies for each sources from strategy S1.}
    \label{fig:regressionS1}
\end{figure}

Figure~\ref{fig:regressionS1} shows the regression lines for each source selected from strategy S1 with \textit{Wang2023} target data. 
That from the strategy S2 has been already shown in Figure~\ref{fig:regressionS2}. 

The numerical values in the linear scaling lines are provided in Table~\ref{tab:regression_params}

\begin{table}
    \caption{Numerical values of the linear scaling for each source.}
    \centering
    \input{Tables/regression_params}
    \label{tab:regression_params}
\end{table}

\begin{figure}
    \centering
    \includegraphics[width=0.8\linewidth]{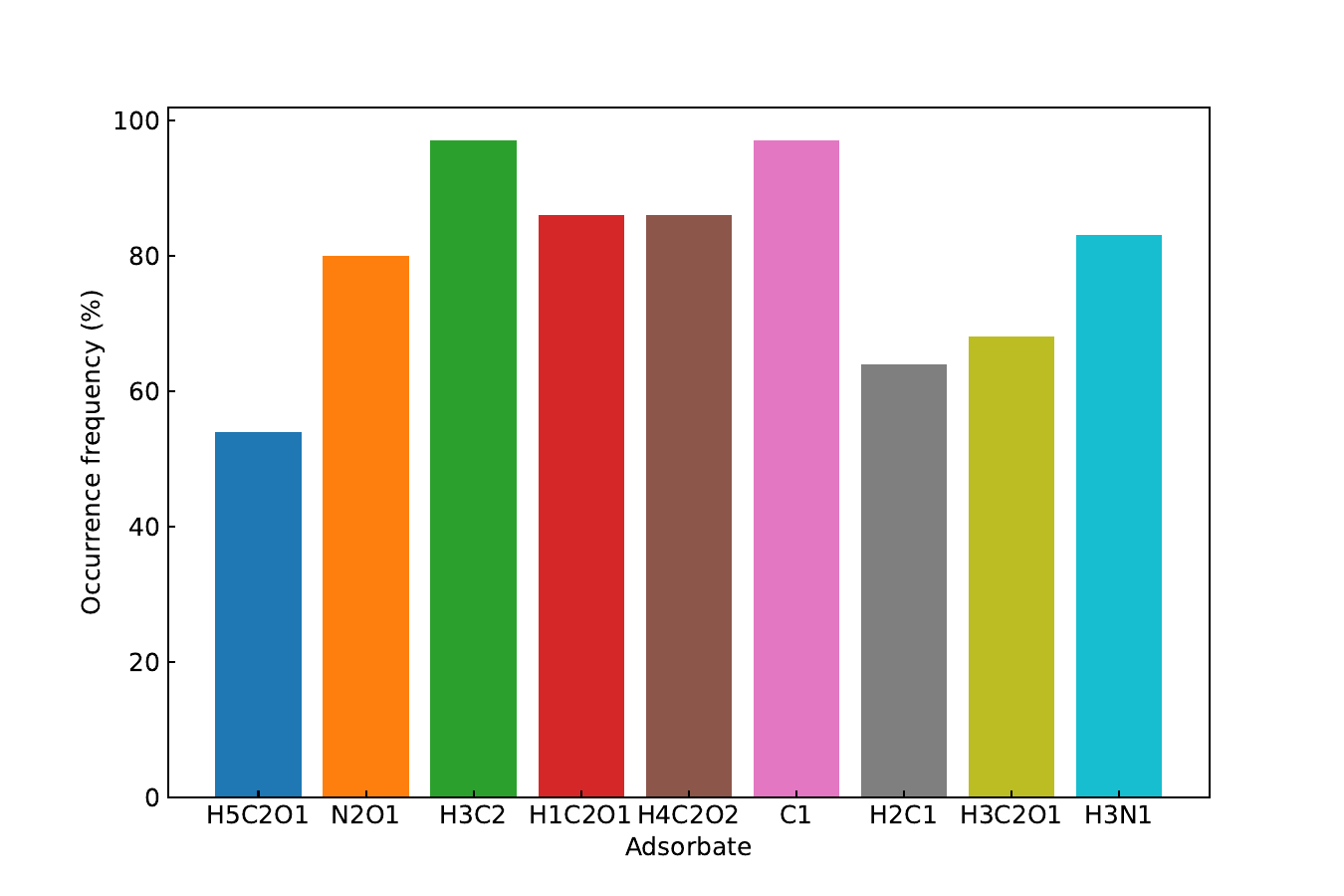}
    \caption{Results of source selection with strategy S2 for dummy data.}
    \label{fig:dummy_imputation}
\end{figure}

While the main text only showed the function estimation for \textit{Wang2023} dataset, here we show the results for the dummy dataset. 
The result of source selection with Algorithm.~\ref{alg:multiple-imputation} for the dummy data is shown in Figure~\ref{fig:dummy_imputation}.

\begin{figure}
    \centering
    \includegraphics[width=0.7\linewidth]{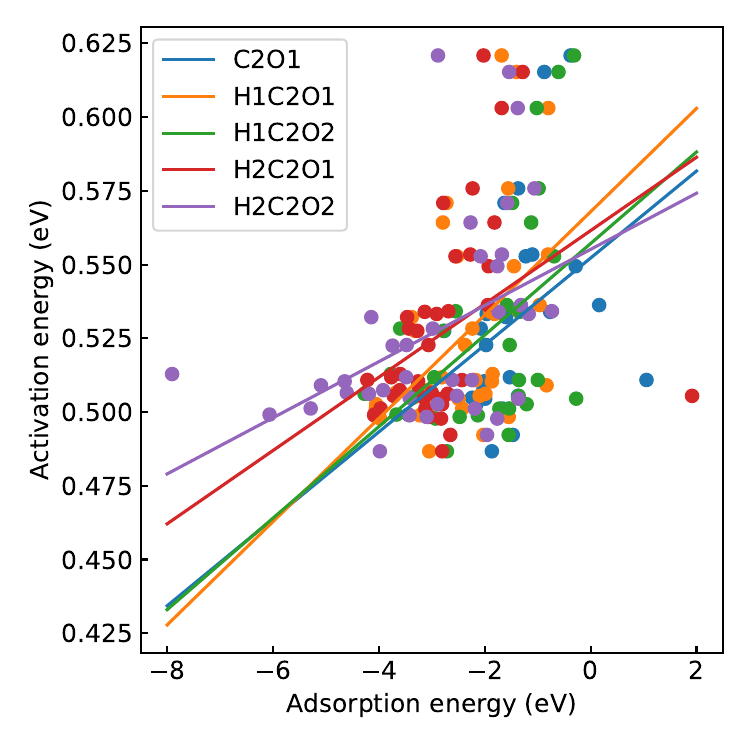}
    \caption{Linear scaling between dummy activation energy and adsorption energies for each sources from strategy S1.}
    \label{fig:dummy_regressionS1}
\end{figure}

\begin{figure}
    \centering
    \includegraphics[width=0.7\linewidth]{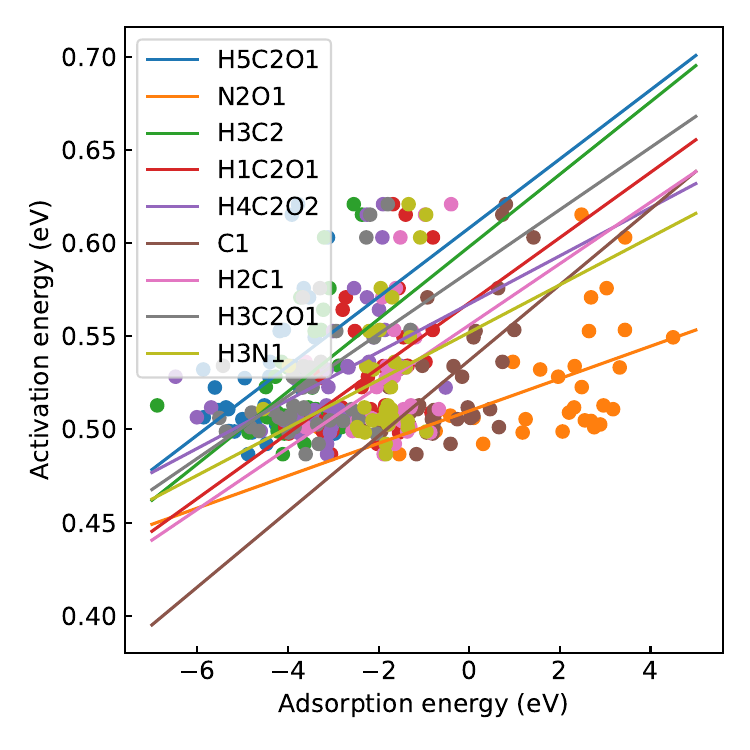}
    \caption{Linear scaling between dummy activation energy and adsorption energies for each sources from strategy S2.}
    \label{fig:dummy_regressionS2}
\end{figure}

The linear scaling lines between the dummy activation energy and the adsorption energies for each strategy are provided in Figures~\ref{fig:dummy_regressionS1} and \ref{fig:dummy_regressionS2}.
The numerical values are shown in Table~\ref{tab:dummy_regression_params}.

\begin{table}
    \caption{Coefficients and intercepts for each regression lines.}
    \centering
    \input{Tables/dummy_regression_params}
    \label{tab:dummy_regression_params}
\end{table}

\subsection{Source training}
\label{sec:app_training}
All essential parameters through our experiments are summarized in Table~\ref{tab:nn-params}.

\begin{table}
    \caption{List of model parameters and experimental conditions of our demonstration.}
    \centering
    \begin{tabular}{lr}
        \toprule
         Parameter & Value \\
         \midrule
        Input dimension & 224 \\
        Output dimension & 1 \\
        Training batch size & 32 \\
        Optimizer & SGD \\
        Learning rate & 0.001 \\
        Random seed (source pretrain) & 0\\
        Random seed (target training) & 0 \\
        Epochs (source pretrain) & 50 \\
        Epochs (target training) & 175 \\
        \bottomrule
    \end{tabular}
    \label{tab:nn-params}
\end{table}

\begin{figure}[ht]
  \centering
  \includegraphics[width=0.7\linewidth]{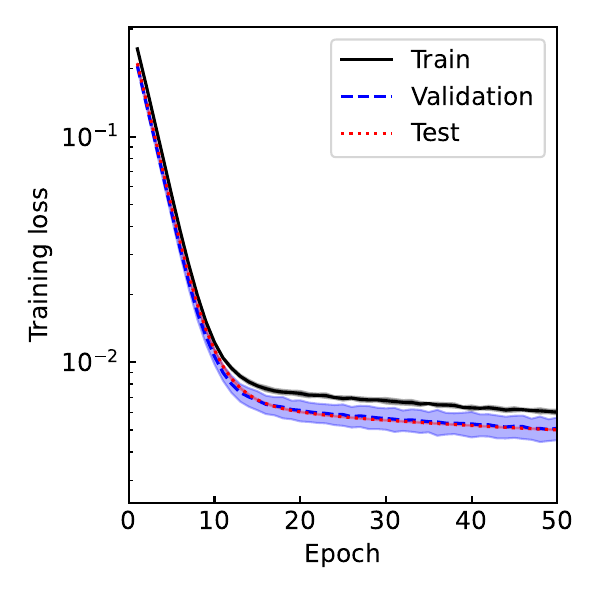}
  \label{fig:sourcetrain}
  \caption{Source training curve for {\sf C2O1} data with respect to the losses of training, validation, and test, as functions of the number of epochs.}
\end{figure}

Figure~\ref{fig:sourcetrain} shows training curves with source data, displaying the mean and the standard deviation of losses obtained by 10-fold cross validation. 
The result of {\sf C2O1} dataset is shown as an example. 
Here, we separated the 20\% of source data for testing, and performed cross validation with the remaining 80\%. 
Apart from the validation, the source model was trained using all the source data, and carried out early stopping on 50 epochs.

\subsection{Target training}
\begin{figure}
  \centering
  \includegraphics[width=0.7\linewidth]{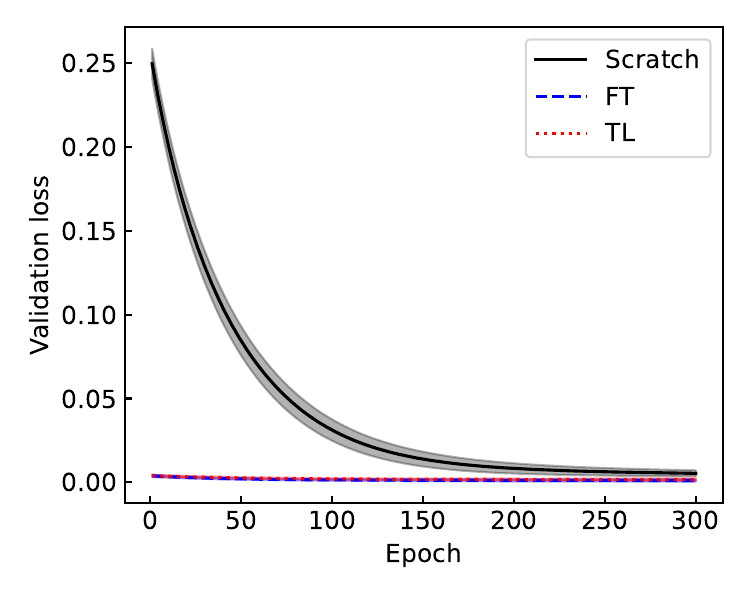}
  \label{fig:targettrain}
  \caption{Target training curve with respect to the validation losses for each training methods, as functions of the number of epochs.}
\end{figure}

Figure~\ref{fig:targettrain} shows the target training curves for each retraining method of the full scratch, the fine tuning, and the transfer learning, explained in Section \ref{sec:model}. 
Here, {\sf C2O1} dataset is used as an example. 
The mean and the standard deviation of validation losses obtained by 10-fold cross validation are displayed.
With the prediction models, we stopped training after 125 epochs.

\begin{figure}[ht]
    \centering
    \includegraphics[width=0.7\linewidth]{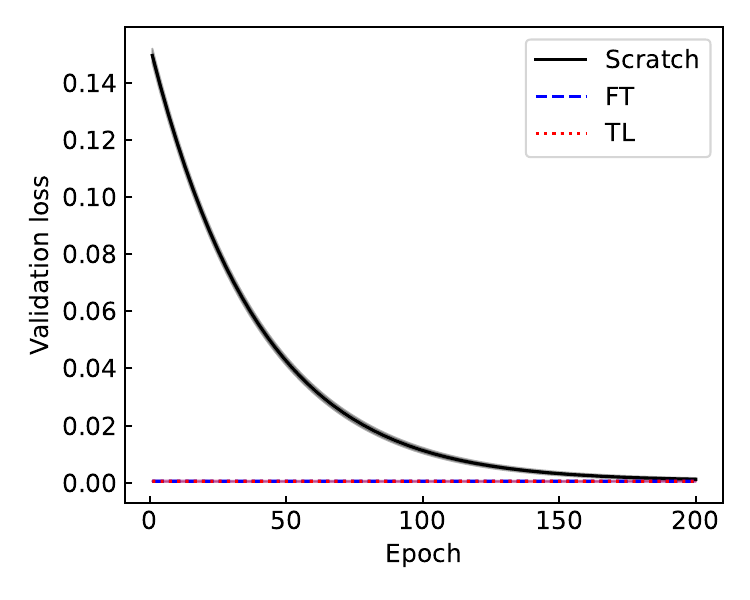}
    \caption{Target training curve with the limited (200) dummy data.}
    \label{fig:dummy_targettrain}
\end{figure}

We also performed 10-fold cross-validation during the target training with the dummy dataset. 
Figure~\ref{fig:dummy_targettrain} shows its target training curve using the {\sf C2O1} source data. 
To compare with the \textit{Wang2023} containing 141 target data, we limit the size of dummy data to 200 instances by random sampling. 
As a result, Figure~\ref{fig:targettrain} and \ref{fig:dummy_targettrain} show similar behavior.

\begin{figure}[ht]
    \centering
    \includegraphics[width=0.7\linewidth]{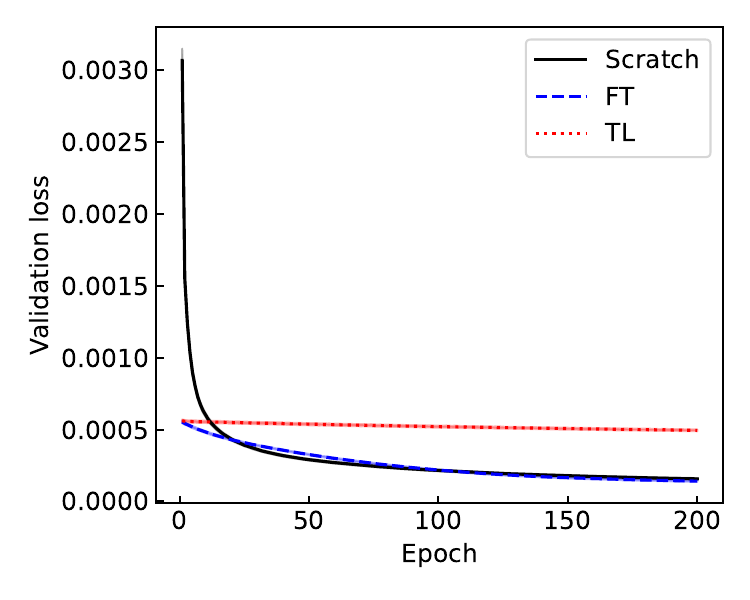}
    \caption{Target training curve with the limited (200) dummy data.}
    \label{fig:dummy_fulltargettrain}
\end{figure}

Using full dummy data, we can investigate the behavior of the target training with a large target dataset, as shown in Figure~\ref{fig:dummy_fulltargettrain}. 
According to the result, the full-scratch and the FT exceed the TL in the case with a large dataset; 
It is reasonable because the last one has a lesser capacity for fixed parameters than the others. 
However, as shown in the main text, there is little difference between FT and TL in the range with O(100) data, and both are superior to the full-scratch model.

\end{document}

%% file: preamble.tex
\newcommand{\calD}{\mathcal{D}}
\newcommand{\calX}{{\mathcal{X}}}
\newcommand{\calY}{{\mathcal{Y}}}

\newcommand{\DD}{\mathbb{D}}
\newcommand{\EE}{\mathbb{E}}
\newcommand{\src}{{\mathrm{S}}}
\newcommand{\tgt}{{\mathrm{T}}}
\newcommand{\dummy}{{\mathrm{T_d}}}

\newcommand{\hypara}{\theta}
\newcommand{\threshold}{\alpha}

\newcommand{\Eads}{E^{\mathrm{ads}}}
\newcommand{\Eact}{E^{\mathrm{act}}}

\newcommand{\Natm}{N_{\mathrm{a}}}
\newcommand{\kB}{k_{\mathrm{B}}}
\newcommand{\chem}[1]{{\mathrm{#1}}}

\newcommand{\TimeOfImputation}{N_{\mathrm{imp}}}
\newcommand{\LASSOlambda}{\lambda}

\newcommand{\numelem}{n_{\mathrm{elem}}}
\newcommand{\maxindex}{l_{\mathrm{max}}}

%% file: Tables/datasize_summary_S1S2.tex
\begin{tabular}{llrrr}
\toprule
Source $M$ & Strategy &  $N_{\src(M)\cap\tgt}$ & Pretrain error (eV) & Final error (eV) \\
\midrule
C2O1 & S1 & 12 & 0.05869 & 0.02088 \\
H3N1 & S2 & 13 & 0.04875 & 0.02253 \\
H1C2O2 & S1 & 11 & 0.02573 & 0.02333 \\
H1C2O1 & S1 & 13 & 0.07309 & 0.02456 \\
H3C1 & S2 & 12 & 0.06917 & 0.02473 \\
H3C2O1 & S2 & 13 & 0.08887 & 0.02498 \\
H1N1O1 & S2 & 13 & 0.06496 & 0.02768 \\
H3C2 & S2 & 13 & 0.10224 & 0.02786 \\
H5C2O2 & S2 & 12 & 0.05854 & 0.02835 \\
H2C2O2 & S1 & 13 & 0.06268 & 0.03262 \\
H2C2O1 & S1 & 13 & 0.06956 & 0.03293 \\
H1O1 & S2 & 10 & 0.04391 & 0.03849 \\
\bottomrule
\end{tabular}

%% file: Tables/regression_params.tex
\begin{tabular}{llrr}
\toprule
Source & Strategy & a & b \\
\midrule
C2O1 & S1 & 0.05016 & 0.64007 \\
H1C2O1 & S1 & 0.04293 & 0.66702 \\
H1C2O2 & S1 & -0.00987 & 0.48225 \\
H2C2O1 & S1 & 0.02107 & 0.61106 \\
H2C2O2 & S1 & 0.02499 & 0.62920 \\
H3N1 & S2 & 0.14285 & 0.78651 \\
H3C2 & S2 & 0.06510 & 0.77389 \\
H1N1O1 & S2 & 0.01590 & 0.55576 \\
H5C2O2 & S2 & 0.02596 & 0.63037 \\
H1O1 & S2 & 0.14461 & 0.66663 \\
H3C1 & S2 & 0.09692 & 0.77077 \\
H3C2O1 & S2 & 0.06735 & 0.79618 \\
\bottomrule
\end{tabular}

%% file: Tables/dummy_regression_params.tex
\begin{tabular}{llrr}
\toprule
Source & Strategy & a & b \\
\midrule
C2O1 & S1 & 0.01474 & 0.55223 \\
H1C2O1 & S1 & 0.01751 & 0.56792 \\
H1C2O2 & S1 & 0.01552 & 0.55714 \\
H2C2O1 & S1 & 0.01243 & 0.56153 \\
H2C2O2 & S1 & 0.00952 & 0.55516 \\
H5C2O1 & S2 & 0.01853 & 0.60810 \\
N2O1 & S2 & 0.00869 & 0.50991 \\
H3C2 & S2 & 0.01944 & 0.59807 \\
H1C2O1 & S2 & 0.01751 & 0.56792 \\
H4C2O2 & S2 & 0.01292 & 0.56735 \\
C1 & S2 & 0.02028 & 0.53705 \\
H2C1 & S2 & 0.01649 & 0.55598 \\
H3C2O1 & S2 & 0.01669 & 0.58458 \\
H3N1 & S2 & 0.01278 & 0.55205 \\
\bottomrule
\end{tabular}